\documentclass[letterpaper, aps, tightenlines, 11pt, nofootinbib, preprintnumbers]{revtex4}
\pdfoutput = 1
\usepackage{graphicx}
 
\usepackage{setspace}
\usepackage[pdftex, hyperfootnotes = true, bookmarks = false, colorlinks = true, linkcolor = blue, citecolor = purple]{hyperref}
\usepackage{titlesec}
\usepackage{xcolor}
\usepackage{amsmath}
\usepackage{amssymb}
\usepackage[percent]{overpic}

\definecolor{Red}{rgb}{1,0,0}

\usepackage[margin = 2.5cm]{geometry}
\def\thesection{\arabic{section}}
\def\thesubsection{\arabic{subsection}}
\titleformat{\section}
  {\normalfont\Large\bfseries}{\thesection}{1em}{}
\titleformat{\subsection}
  {\normalfont\large\it}{\thesection.\thesubsection}{1em}{}
\titleformat{\subsubsection}
  {\normalfont\sl}{\thesection.\thesubsection.\thesubsubsection}{1em}{}

    \newcommand\be{\begin{equation}}
\newcommand\ee{\end{equation}}

  \newcommand\ba{\begin{aligned}}
\newcommand\ea{\end{aligned}}

\newcommand{\bbb}{\ba\begin{array}{c}}
\newcommand{\eee}{\nonumber\end{array}\ea}
\newcommand{\een}[1]{\label{#1}\end{array}\ea}

\usepackage{xcolor}
\definecolor{Red}{rgb}{1,0,0}
\definecolor{Blue}{rgb}{0,0,1}

\renewcommand\Im{ {\rm Im}\,}
\renewcommand\Re{ {\rm Re}\,}

  \newcommand{\IR}{\relax{\rm I\kern-.18em R}}

\def\D{\Delta}
\def\bD{{\bar \Delta}}
\def\bq{{\bar q}}

\def\Z{\mathbb{Z}}
\def\R{\mathbb{R}}

\def\H{\mathbb{H}}
\def\C{\mathbb{C}}

\def\LP{\mathcal{L}}

\def\cA{\mathcal{A}}
\def\cC{\mathcal{C}}
\def\cB{\mathcal{B}}
\def\cE{\mathcal{E}}
\def\cL{\mathcal{L}}
\def\cP{\mathcal{P}}

\def\ie{{\it i.e.~}}
\newcommand{\eg}{{\it e.g.~}}

\def\j{j}
\def\e{e}
\def\J{J}
\def\E{E}

\def\sh{{\xi}}

\def\Zp{Z^p}
\def\Zc{F^p}

\def\SL{SL(2,\Z)}
\def\Gcusp{\Gamma_\infty}

\def\Laplace{\nabla^2}

\def\ZpureP{Z_{pure}^P}
\def\Zpure{Z_{pure}}
\def\Zpurep{Z_{pure}'}

\begin{document}

\begin{titlepage}

\begin{center}
{\Large \bf Poincar\'e Series, 3D Gravity and CFT Spectroscopy}

\vspace*{6mm}

Christoph A.~Keller$^a$ and Alexander Maloney$^{b,c}$

\vspace{5mm}

\textit{ 
$^a$ NHETC, Rutgers, The State University of New Jersey, 
Piscataway, 
USA}\\ 

\vspace{2mm}

\bigskip
\textit{$^b$ Department of Physics, McGill University, Montreal, Canada}
\vspace{2mm}

\bigskip
\textit{$^c$ Center for the Fundamental Laws of Nature, Harvard University, Cambridge, 
USA}
 
\vspace{5mm}

{\tt  keller@physics.rutgers.edu, maloney@physics.mcgill.ca  }

\vspace*{1cm}
\end{center}

\begin{center}
{\bf Abstract}
\end{center}

 Modular invariance strongly constrains the spectrum of
states of two dimensional conformal field theories.
By summing over the images of the modular group, we construct
candidate CFT partition functions
 that are modular invariant and have positive spectrum.  This allows
us to efficiently
 extract the constraints on the CFT spectrum imposed by modular
invariance, giving information on the spectrum
that goes beyond the Cardy growth of the asymptotic density of states.
 Some of the candidate modular invariant partition functions we
construct have gaps of size $(c-1)/12$, proving that gaps of this size
and smaller are consistent with modular invariance.
We also revisit the partition function of pure Einstein gravity in
AdS$_3$ obtained by summing over geometries, which has a spectrum with
two unphysical features: it is continuous, and the density of states
is not positive definite. We show that both of these can be resolved by adding corrections to the spectrum which
are subleading in the semi-classical (large central charge) limit.

\vspace{1cm}

\begin{flushleft}{\small RUNHETC-2014-13}\end{flushleft}

\end{titlepage}

\section{Introduction and Summary}

\subsection{Modular invariance in CFT}

For conformal field theories in two dimensions, 
modular invariance -- the invariance under large conformal transformations in Euclidean signature -- strongly constrains the spectrum of the theory. 
Famously, Cardy showed that it determines the asymptotic density of states at high energy universally \cite{Cardy:1986ie}. 
In later work, modular invariance was used to obtain subleading corrections to this behavior \cite{Carlip:2000nv},
information about states of intermediate energy \cite{Hellerman:2009bu,Qualls:2013eha}, and
the phase diagram of the free energy \cite{Hartman:2014oaa}.  
What all these results have in common is that they only use invariance under a single element of the modular group -- S duality -- which states that the finite temperature partition function is invariant under $T\to 1/T$.    
In this paper we will study more generally the constraints placed by invariance under the full modular group.  We will do so by understanding better the structure of the space of non-holomorphic modular-invariant functions, using a method inspired by the AdS$_3$/CFT$_2$ correspondence.

The partition function of a CFT$_2$ is
\begin{equation}
Z(\tau, \bar \tau) = \sum q^{h-c/24} {\bar q}^{\bar h-c/24},~~~~~~q=e^{2 \pi i \tau}
\end{equation}
where the sum is over all states in the spectrum, and $(h,\bar h)$ are the left- and right-moving conformal dimensions.  
These dimensions are normalized so that the vacuum state has $(h,\bar h)=(0,0)$.
Since $h+\bar h $ and $h - \bar h$ are the energy and angular momentum of the state, respectively, $\Im \tau$ can be regarded as the inverse temperature and $\Re \tau$ as a thermodynamic potential associated with angular momentum. 
The statement of modular invariance is that 
\be
Z(\tau, \bar\tau) = Z(\tau|_\gamma,\bar\tau|_\gamma)
\ee
for any element $\gamma = \left({a~b\atop c~d}\right) \in \SL$, where 
\be
 \tau|_\gamma = \frac{a\tau+b}{c\tau+d} \,.
\ee 
This follows from the fact that $Z(\tau,\bar \tau)$ can be interpreted as the partition function of the CFT on the torus 
$z\sim z+1\sim z+\tau$, whose conformal structure is invariant under $\tau\to\tau|_\gamma$.

We will find it convenient to write the partition function as 
\be
Z(\tau, \bar \tau) = \int_0^\infty dh d\bar h~ \rho(h,\bar h)~  q^{h-c/24} {\bar q}^{\bar h-c/24}
\ee
where $\rho(h, \bar h)$ is a spectral density.  Modular invariance then translates into a set of constraints on $\rho(h,\bar h)$.
For a CFT with a discrete spectrum $\rho(h,\bar h)$ is a sum of delta functions, but it is often useful to approximate $\rho(h,\bar h)$ by a continuous density of states.  
The goal of the present paper is to study the space of modular invariant functions $Z(\tau,\bar \tau)$ and their corresponding density of states $\rho(h,\bar h)$. 

\subsection{$AdS_3$/CFT$_2$}

The holographic correspondence \cite{Maldacena:1997re}
relates two dimensional conformal field theories to three dimensional theories of gravity in AdS$_3$.
In this correspondence the CFT central charge is\be
c = 3l/2G
\ee
where $l$ is the $AdS$ radius and $G$ is Newton's constant.
It is natural to ask what modular invariance corresponds to
on the gravity side.  The authors of \cite{Dijkgraaf:2000fq} proposed the following:
modular invariance arises from the sum over saddle points
of a gravitational path integral. 
In particular, one can compute the torus partition function of the CFT by summing over three dimensional Euclidean geometries whose asymptotic boundary is a torus.
One such geometry is Euclidean AdS$_3$ with the Euclidean time direction periodically identified; this is the ``thermal AdS" geometry describing a finite temperature ensemble in AdS$_3$. Another such geometry is the Euclidean BTZ black hole \cite{Banados:1992wn}. 
There are in fact an infinite number of such geometries, each labelled by an element of the modular group $\SL$\footnote{More precisely, the group in question is a subgroup of $\SL$. We will make this statement more precise in section~\ref{s:Zp}.}\cite{Maldacena:1998bw}.
The path integral therefore includes a sum over the modular group $\SL$, which renders the partition function modular invariant.

From the CFT point of view, one way of understanding this sum is to start with the contribution\be 
q^{h-c/24} \bq^{\bar h - c/24}
\label{orig}
\ee 
to the partition function of a state with dimension $(h, \bar h)$.  On its own, this contribution is not modular invariant.
However, the sum over $\SL$
\be\label{poin}
\sum_{\gamma\in\SL} q^{h-c/24} \bar q^{\bar h -c/24}|_{\gamma}
\ee 
is modular invariant, provided that the sum makes sense.  Expressions like (\ref{poin}) are known as {\it Poincar\'e series}.
Starting with the original state (\ref{orig}), the non-trivial $\SL$ images in the sum (\ref{poin}) will lead to new states in the spectrum.  
We seek to understand these new states.  

In the simplest case, one starts with the contribution $|q|^{-c/12} = \exp\left\{ c {\pi \over 3} \Im \tau\right\}$ of the vacuum state. In the gravitational language, this is interpreted as the semi-classical contribution to the partition function of thermal AdS$_3$; with our normalization, empty AdS$_3$ has energy $-c/6$, and Euclidean time is periodically identified with period $2\pi~ \Im \tau $.  
The sum over geometries then leads to the Poincar\'e series (\ref{poin}) with $h =\bar h=0$. 
We would like to interpret the new states arising from the $\SL$ sum as black hole states.
To begin, let us recall that 
a BTZ black hole of mass $M$ and spin $J$
can be interpreted as a CFT state with (see \eg \cite{Kraus:2006wn})
\be
h-c/24 =\frac{1}{2}(Ml-J), \qquad \bar h-c/24 = \frac{1}{2}(Ml+J)\ .
\label{bhmj}
\ee
The black hole will have a smooth horizon only if it satisfies the cosmic censorship condition $|J| \le Ml$.
So a state can be interpreted as a black hole only if $h$ and $\bar h$ are both greater than $c/24$.  We will show that -- up a a subtlety discussed below -- the new states coming from the $\SL$ sum do indeed have this property.

One important subtlety is that we only wish to interpret primary states as black holes.  Descendant states are interpreted as perturbative excitations built out of non-trivial diffeomorphisms -- known as boundary gravitons -- applied to a primary state, which could either be the vacuum or a black hole state. 
This will modify the above statements somewhat.
Indeed, each primary state will be dressed by an infinite tower of descendant states, which must be added to the contribution (\ref{orig}) of that state to the partition function.  Including these states, a  primary of dimension $(h, \bar h)$ will give a contribution\be
q^{\Delta}{\bar q}^{\bD} |\eta(\tau)|^2
\ee
to the partition function.  Here 
\be
\Delta = h - \sh,~~~~ \bD = \bar h - \sh,~~~~~\sh=\frac{c-1}{24}
\ee
are the shifted dimensions which include an additional contribution coming from descendants.  
In fact, we will argue below that it is the states with $\D, \bD$ positive which should be interpreted as black holes, rather than those with $h-c/24$ and $\bar h-c/24$ positive.
Indeed, equation (\ref{bhmj}) is valid only in the semi-classical (large $c$) limit, so a correction of this form is expected due to one-loop effects.  So this discrepancy can be interpreted as a shift of the renormalized mass of the lightest BTZ black hole at one-loop.

\begin{figure}[htbp]
\begin{center}
\includegraphics[height=.3\textwidth]{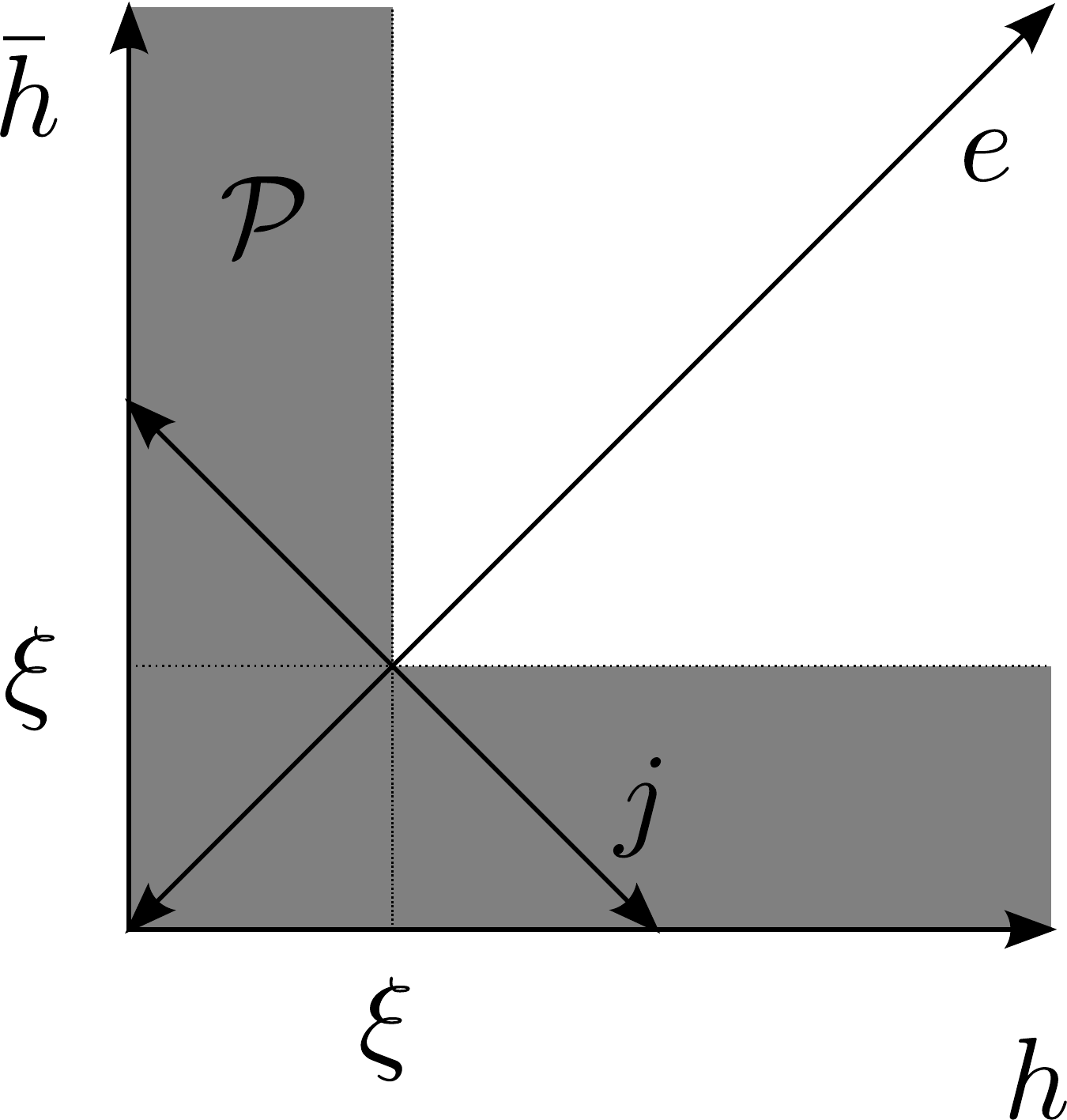}
\caption{The censored region $\cP$ with $\sh=\frac{c-1}{24}$,
and the shifted energy and spin $e=\D+\bD$, $j=\D-\bD$.}
\label{Fig:censored}
\end{center}
\end{figure}
To state our results more precisely, let us define a \emph{censored} state as one
contained in the set
\be
\cP = \{ (h,\bar h) : h<\sh\ \textrm{or}\ \bar h < \sh \}\ .
\ee
These are the states which, in the gravitational language, cannot be interpreted as BTZ black holes. 
These states will play a special
role for modular invariant partition functions,
somewhat similar to that played by polar states in the theory of modular forms or weak Jacobi 
forms. We will not call such states polar though,
as this name is more natural for states with $h+\bar h < \frac{c}{12}$ (i.e. those states which
give a divergent contribution as $\tau \rightarrow i\infty$).  States which are not in $\cP$ will be called \emph{uncensored}.

One of our main results is that the Poincar\'e images will give a contribution to $\rho(h, \bar h)$ which lies in the uncensored region. 
Thus we can really interpret them as black holes.
On a formal level this is  easy to see. As we will explain in section~\ref{s:Zp}, the
Poincar\'e series (\ref{poin}) is
\be
\sum_{\gamma: c\geq0,(c,d)=1} q^{h-c/24} \bar q^{\bar h -c/24}|_{\gamma}\ . 
\ee
Consider this as a function of two independent complex variables $\tau$ and $\bar \tau$.
Since the only $\gamma$ in the sum with $c=0$ is the identity element,
the exponent remains finite in the limit $\tau \rightarrow i\infty$ 
for all images in the sum.  This means
that all of the new states which appear must have $h-c/24 \ge 0$.  Once the contributions of the descendants are included, this becomes $\Delta \ge 0$. 
The same argument holds of course for $\bar \tau$ and $\bar h$. The problem
with this argument is that Poincar\'e series (\ref{poin}) is divergent
and needs to be regularized. Regulating such a sum 
is quite subtle and in some cases can 
change properties that one would naively expect (see e.g. \cite{Manschot:2007ha}).
We will discuss the regularization in detail and show that it does not change this basic property.

\subsection{Partition functions and free energy}

We can now state our main result. Given a primary state
of weight $(\D,\bD)$, we construct a partition function
$Z_{\D,\bD}(\tau)$ with the following properties:
\begin{itemize}
\item $Z_{\D,\bD}(\tau)$ is invariant under $\SL$.
\item The spectrum $\rho(h,\bar h)$ of $Z_{\D,\bD}(\tau)$ is a continuous function of the energy $h+\bar h$, and delta function supported at integer values of the angular momentum $h-\bar h$.
\item If $(\D,\bD)$ is uncensored, then the spectrum $Z_{\D,\bD}(\tau)$
has no censored states. If $(\D,\bD)$ is censored, then
it is the only censored primary state in the spectrum.
\item If $(\D,\bD)$ is censored, $-(\D +\bD)$ is large enough,
and $|\D-\bD|$ is not too large, then the density of states
in the spectrum is positive.
\item If instead of a primary field we take the vacuum, then the spectral density $\rho(h, \bar h)$ is continuous and the identity is the only censored primary state. Moreover, at large $c$ the density $\rho(h, \bar h)$ is positive with the exception of an $O(1)$ number of states with $\Delta=\bD=0$.
\end{itemize}
We obtain $Z_{\D,\bD}(\tau)$ by computing the Poincar\'e series (\ref{poin}) explicitly.
We will use a version of the construction of \cite{Maloney:2007ud}, modified slightly to ensure that
the density of states is positive.

From these properties various results follow. First, note that given
any censored spectrum, by a linear combination of the above results
we can always obtain a modular invariant function with that particular censored spectrum. 
In this sense our results are an existence proof. 
They are not, however,
a uniqueness result: we will argue that in general there
are a great many modular invariant functions with a given
censored spectrum. 

It is useful to compare this to the case of holomorphic modular functions, which would be relevant if we were 
studying the partition function of a chiral CFT or the elliptic genus of an ${\cal N}=2$ SCFT.
As far as existence is concerned, the censored region is very similar to the
polar region in the case of holomorphic modular functions:
for any given choice of polar states, there is
always a modular invariant function which has this polar
spectrum. 
For uniqueness, however, the situation is completely
different: in the holomorphic case, the polar part (in which
we also include the constant term for convenience)
fixes the modular function completely.
For non-holomorphic modular functions this is not the case. To put it another way, there
are a great many non-holomorphic modular functions whose
censored spectrum vanishes. A very simple example which will
play a role later on is the Eisenstein series
\be
E(\tau,s) = \sum_\gamma y^s|_\gamma\ .
\ee

Of course, if we want to interpret our modular functions
as partition functions of physical theories, we
must also demand that the density of states is
positive. Ensuring this is more subtle, and there is
no reason to believe that an arbitrary censored configuration
will give a positive spectrum. In general, however,
if there are not too many censored states of high spin,
we will show that the density of states is indeed positive. In particular
for diagonal theories  -- theories that only
have scalar censored primary fields -- the density
of states will be positive.

Our methods also allow us to determine certain features of the free energy from the censored part of the spectrum.
Let us begin by considering the holomorphic case, where the free energy can be determined exactly from the polar part of the spectrum.
The partition function $Z(\tau)$ is a meromorphic function on
the quotient $\H/\SL$ whose only pole is at $\tau =i\infty$.
The polar part of the partition function takes the form
$Z_{pol}(\tau) = \sum_{k=1}^{c/24} a_{-k}q^{-k}$ for some constants $a_{-k}$.
To turn this polar part into a modular invariant function,
we perform the holomorphic version of the Poincar\'e series, known as a Rademacher sum. The function $\tilde Z_{pol}(\tau)$ so obtained
still has polar part $Z_{pol}(\tau)$.
It follows that $Z(\tau)-\tilde Z_{pol}(\tau)$
is a bounded, holomorphic function on the compact space
$\H/\SL$, hence it is
a constant.  Thus the free energy is determined exactly from the Rademacher sum of the polar part.

In the non-holomorphic case we will proceed along similar lines, constructing the Poincar\'e series of the censored
part of $Z(\tau, \bar \tau)$.  This Poincar\'e series will agree with the original $Z(\tau, \bar \tau)$ up to a function which is bounded on $\H/\SL$. However, because the function is not holomorphic, it is not necessarily constant. In fact the space of bounded
modular functions is infinite dimensional.  Nevertheless, we can
still  use the Poincar\'e series to determine
the free energy up to a function which is bounded as $\tau\to i\infty$. In other words, the Poincar\'e series determines the free energy
up to a finite piece.

\subsection{Pure Gravity}

These results have interesting implications for the potential existence of ``pure" theories of quantum gravity in AdS$_3$, i.e. theories which contain only a metric and no other degrees of freedom.  In \cite{Maloney:2007ud} the partition function of pure Einstein gravity was, under certain plausible assumptions, shown to be precisely the Poincar\'e series starting with the vacuum state described above. 
It was further argued that the resulting partition function does not have a sensible quantum mechanical interpretation, as it cannot be interpreted as the trace over a discrete Hilbert space with positive norm.  We can now refine this result, and show that the resulting partition function is nonsensical in precisely two ways: 
\begin{itemize}
\item
The spectrum $\rho(h, \bar h)$ is continuous.
\item
The spectrum $\rho(h, \bar h)$ is not positive definite.
\end{itemize} 
We will argue that both of these problems can be fixed by adding a correction to the partition function which is subleading in the large central charge limit.  In particular, this new correction term can be interpreted as an intrinsically quantum mechanical contribution to the partition function which is invisible in the semi-classical limit.  While we are not able to give a bulk interpretation for this additional contribution, this may suggest that a small modification of the gravitational path integral could give a sensible quantization of Einstein gravity in three dimensions.

\subsection{Modular bootstrap and gaps}
Our results  connect to the
conformal bootstrap program for the partition
function started in \cite{Hellerman:2009bu} and 
continued in \cite{Keller:2012mr,Friedan:2013cba}.
The ultimate goal of this program is to classify all modular invariant
partition functions that could come from
2d CFTs. This would give all possible
CFT spectra. One important feature of the spectrum is the size of the gap, i.e.
the conformal weight $\Delta_1:=h+\bar{h}$
of the lowest lying non-vacuum primary.
In \cite{Hellerman:2009bu}, S-invariance was used to bound the gap as a function of the
central charge:\be
\Delta_1 < \frac{c}{6}+0.474\ .
\ee
In
\cite{Friedan:2013cba} this bound was improved in
a systematic way.  
For small $c$ these bootstrap methods converged rapidly, 
but for large $c$ the problem becomes numerically more difficult. 
One would like to find the strongest possible 
bound on $\Delta_1$, or at least to obtain a lower bound for the bound. 
In holomorphically factorized theories,
the product of two extremal partition functions
(as defined in \cite{Witten:2007kt}) gives a partition
function whose lowest primary has
$\Delta_1=\frac{c}{24}+1$. Holomorphic
factorization is a very strong constraint,
so we expect that general theories might have
larger gaps.

Our results give a lower
bound for the bound: by constructing
explicit examples of partition functions with
$\Delta_1 = 2\sh$, we show that 
no bound derived from modular invariance alone can be stronger than that. 
In \cite{Friedan:2012jk}
it was already argued that no stronger bound than $2\sh$
can be obtained by requiring the partition function
$Z(\tau)$ to be invariant under $S$. 
Our results imply that
imposing full $\SL$ invariance 
cannot improve on the situation.

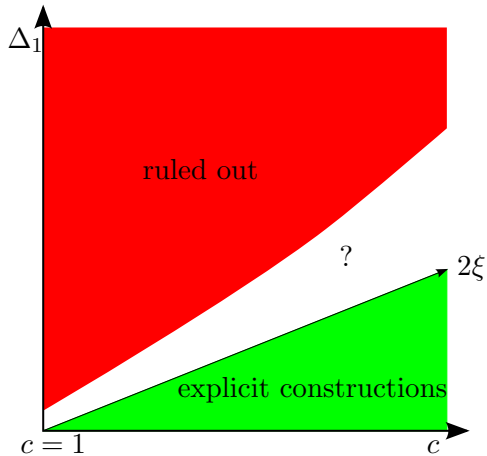
\begin{figure}[htbp]
\begin{center}
\begin{overpic}[height=.35\textwidth]{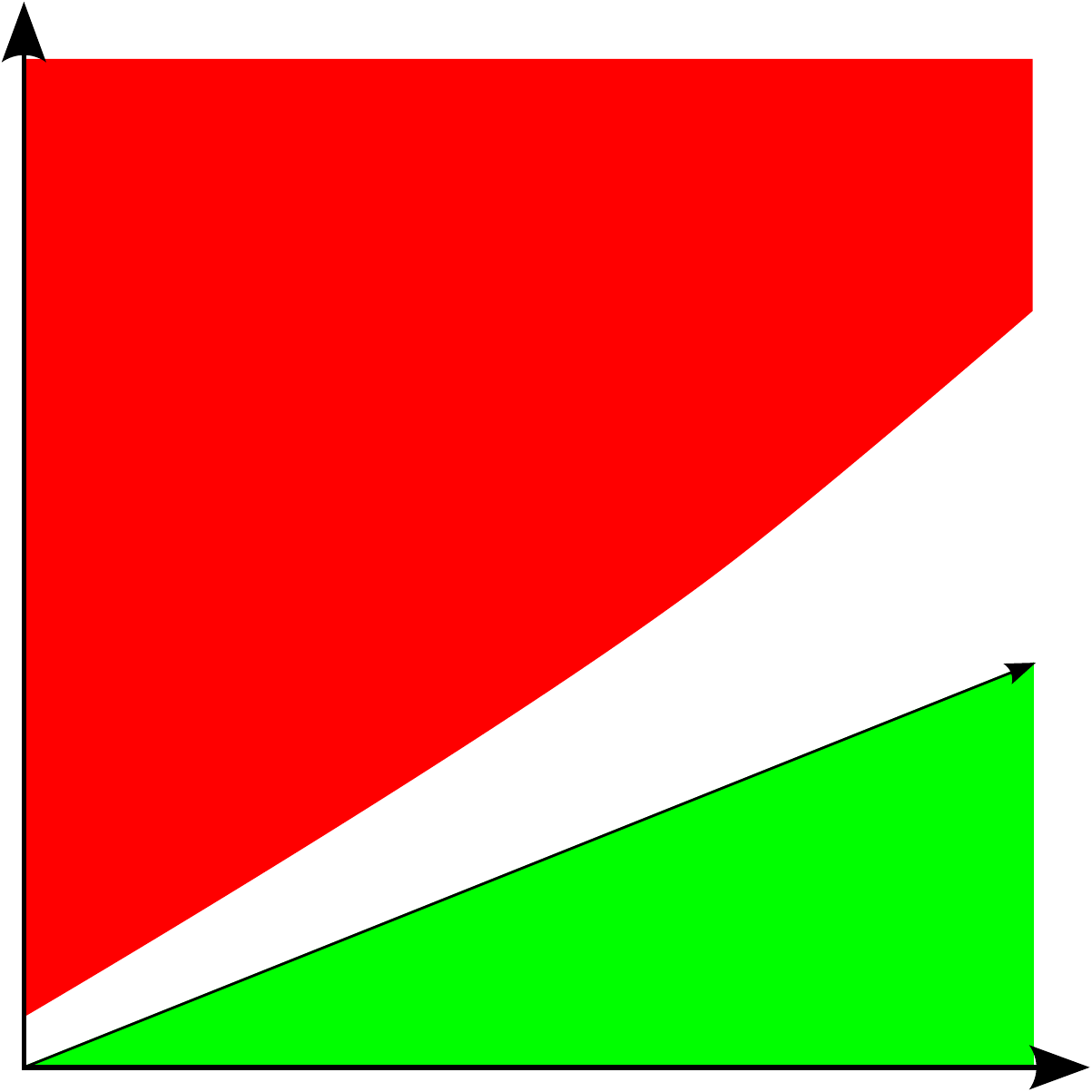}
\put(-6,90){$\Delta_1$}
\put(90,-3){$c$}
\put(-3,-3){$c=1$}
\put(25,60){ruled out}
\put(70,40){?}
\put(33,10){explicit constructions}
\put(97,38){$2\sh$}
\end{overpic}
\caption{The space of modular invariant
partition functions, plotting the conformal weight $\Delta_1$
of the lowest primary against the central charge $c$ of the theory. 
The red region is ruled out by conformal bootstrap methods.
For the green region we construct explicit examples of partition functions.
The status of the white wedge between is still an open question.}
\label{Fig:gap}
\end{center}
\end{figure}

\subsection{Summary}
Since the full Poincar\'e series is somewhat
technical, we will begin by discussing a finite baby version of the sum
in section~\ref{s:warmup}. This avoids all issues
related to regularization, but still exhibits 
many of the most important features of the full sum. We also
use the opportunity to discuss how our results relate
to the bootstrap program. In section~\ref{s:Zp} 
we extend the analysis of \cite{Maloney:2007ud}
to compute explicit expressions for the Poincar\'e 
series. 
In section~\ref{s:spectrum} we compute the inverse
Laplace transform of those expressions to
obtain the spectrum, and show that it satisfies
the properties listed above. In section~\ref{s:gravity}
we discuss in more detail the implications of our results
for the existence of pure gravity in $AdS_3$.

\section{Warmup: self-reciprocal functions}\label{s:warmup}

As a warmup, rather than considering the full modular invariance of the partition function, we will consider the invariance under the $S$ transformation $S:\tau \to -1/\tau$.   Functions which are invariant
under  $S$ are sometimes called
self-reciprocal functions. In this case  
the ``sum over images" is finite (having only two terms) so there are no issues
of regularization. It turns out that most of
the structure we find is the same as for the full
Poincar\'e series. 

\subsection{The Cardy contribution}

To make our computations a bit more specific, we consider a CFT which does not possess an extended chiral algebra, and assume that the Virasoro representations do not contain null states.  This is the generic situation for a CFT with $c>1$.  
A primary state of dimension $(h,\bar h)$, along with all of its descendants, will give a contribution to the partition function of the form\footnote{For brevity we will typically denote the partition function $Z(\tau, \bar \tau)$ as $Z(\tau)$, even though it is not (unless otherwise indicated) necessarily a holomorphic function of $\tau$.}
\be
Z(\tau)= \dots + q^{h-\sh} \bar q^{\bar h-\sh}|\eta(\tau)|^{-2} + \dots
\label{contrib}
\ee
Here we have used the fact that the $L_n$ descendants are enumerated by the infinite product
$\prod_n(1-q^n)^{-1} = q^{-1/24} \eta (\tau)$.
Inspired by (\ref{contrib}), we define the partition function of primaries, $\Zp$, by
\be
\Zp(\tau) = Z(\tau) y^{1/2 }|\eta(\tau)|^2
\ee
where $\tau=x+iy$. $\Zp$ counts the number of primary states in the theory.
For convenience we define $\Delta = h - \sh$ and
$\bD = \bar h -\sh$, so that $Z^p(\tau)$ is
\be 
Z^p(\tau) = y^{1/2}\left( q^{-\sh}\bq^{-\sh} |1-q|^2 + \sum_{primaries} q^{\D}\bq^{\bD} \right)\ .
\ee
In this expression we have separated out  the contribution of the vacuum, which is $SL(2,\R)$ invariant and hence annihilated by $L_{-1}$ (leading to the factor of $|1-q|^2$), from those of the other primaries.  
Since $y^{1/2} |\eta(\tau)|^2$ and $Z(\tau)$ are both modular invariant, $Z^p(\tau)$ is modular invariant as well.
The contribution
to $\Zp$ of 
a primary $(h,\bar h)$ is 
\be\label{Zp}
\Zc_{\Delta,\bD}(\tau) = y^{1/2} q^{\Delta} \bar q^{\bD}\ .
\ee

The basic $S$-invariant function constructed from 
$\Zc_{\Delta,\bD}$ is then given by the sum over images:
\be\label{Ssum}
\Zp_{\D,\bD}(\tau)=\Zc_{\Delta,\bD}(\tau) + \Zc_{\Delta,\bD}(-1/\tau)
= \Zc_{\Delta,\bD}(\tau) + \int_0^{\infty} d\Delta' d\bD' 
\rho_{\Delta,\bD}(\Delta',\bD') \Zc_{\Delta',\bD'}(\tau)\ .
\ee
Here $\rho_{\Delta,\bD}$ is the density of states 
coming from the image. As we will see below,
the support of $\rho_{\Delta, \bD}$ is indeed in $\R_+\times\R_+$.  In other words,
the image only contains uncensored states.
To compute $\rho_{\Delta,\bD}$ explicitly, introduce new variables
$(u_1,u_2):= (\sqrt{2\Delta},\sqrt{2\bD})$
and define
\be
f_\tau(\vec u) = y^{1/2} e^{\pi i \tau u_1^2} e^{-\pi i \bar \tau u_2^2} = \Zc_{\Delta,\bD}(\tau)\ .
\ee
We  claim that $f_{-1/\tau}$ is simply
the two dimensional Fourier transform of $f_\tau$.
This is straightforward to check using the 
Gaussian integral
\begin{multline}
\int d^2 u f_\tau(\vec u) e^{2\pi i \vec{u}\cdot\vec{v}}
= y^{1/2} \int du_1 e^{\pi i \tau (u_1^2+2u_1v_1/\tau + v_1^2/\tau^2)} e^{-\pi iv_1^2/\tau}
\times (v_1\rightarrow v_2)\\
= y^{1/2} |i\tau|^{-1} e^{-\pi iv_1^2/\tau} e^{\pi i v_2^2/\bar{\tau}}
= f_{-1/\tau}(\vec v)\ .
\end{multline}
Note that this argument works regardless of whether $\vec v$ is
real or imaginary, since the integral converges in both cases.
Moreover the integral is over real $\vec u$ in both cases,
so that the density is non-vanishing only for $\Delta',\bD'>0$.
We find
\be
\rho_{\Delta,\bD}(\Delta',\bD') = \frac{1}{4}|u_1u_2|g(u_1,u_2)
\ee
where for convenience we choose $g$ to be even, 
\be \label{gC}
g(u_1,u_2) = \cosh(2\pi iu_1v_1)\cosh (2\pi iu_2v_2)\ .
\ee
If we choose $\Delta,\bD <0$, then $\rho_{\Delta,\bD}$
is positive, otherwise it oscillates. 

At first sight, it is tempting to identify 
\be
\label{candidate}
\sum_{(\Delta, \bD) \le 0} \Zp_{\D,\bD}(\tau)
\ee 
as the full partition function of the theory;  the sum here is only over those states in the censored region ${\cal P}$.  
This function is $S$-invariant, and the censored part of its spectrum agrees with that of the full partition function.  The uncensored part of its spectrum is then fully determined by $S$-invariance.
Unfortunately, (\ref{candidate}) can not tell the full story.  First, it is not invariant under the full modular group. We will address this in the next section by modifying $ \Zp_{\D,\bD}(\tau)$ to include the full sum over $\SL$ images, rather than just $S$.
More importantly, $\rho_{\Delta, \bD}$ describes a continuous, rather than discrete, spectrum.  
Thus, by itself (\ref{candidate}) does not describe a sensible CFT spectrum.
Indeed, we will see that (\ref{candidate}) is not the only possible partition function whose censored part matches
that of the full CFT: there are many different partition functions 
whose censored parts agree, but whose uncensored parts
differ.  Thus in the full CFT partition function $\rho_{\Delta, \bD}$ must be augmented by additional terms which render the spectrum discrete.\footnote{The only exception to this is the case of a CFT which is fully holomorphically factorized, in which case the sum over images becomes a Rademacher sum, which gives a discrete spectrum.  This will be reviewed in more detail below.}

\subsection{Non-uniqueness}

We will now show that there are non-vanishing
$S$-invariant functions whose censored spectrum
vanishes.  This means that, in particular, the simple inclusion of $S$-images of the censored states does not completely determine the partition function.

A simple example is obtained by repeating the above analysis for a state
with $\Delta,\bD >0$.
More generally, take a partition function
with an even spectrum $g(\vec u)$.
We can then compute its $S$-transform: 
\be
Z(-1/\tau) = 
\int_{\R^2} d^2u g(\vec u) \int d^2 v f_\tau(\vec v) e^{2\pi i \vec{u}\cdot\vec{v}}
= \int_{\R^2} d^2v f_\tau(\vec v) \hat{g}(\vec v) 
\ee
where $\hat{g}$ is the Fourier transform of $g$.
To get $Z(\tau) = Z(-1/\tau)$ we can require
\be \label{Finv}
g =\hat g\ .
\ee
It follows that any even function in two variables that
is invariant under Fourier transformation gives a 
modular invariant function with positive support.

This computation may at first seem paradoxical, since 
it seems to imply that the modular transform of
both censored and uncensored states only ever gives 
an uncensored spectrum. This would clearly contradict
the observation that there are invariant
partition functions with censored states. The resolution
is that we have implicitly assumed that the Fourier
transform of $g$ exists, which is 
not the case if, for instance, $g$ is not bounded. The resolution
is that the modular transform of a censored state 
gives a density of states which does not have
a Fourier transformation, as can be seen
from (\ref{gC}) directly.

This shows that the classification of
modular invariant functions can be rather subtle.
Naively one could argue that 
the above computation  tells us that invariant
uncensored functions are in one-to-one correspondence
with Fourier-invariant functions. There is of course
a well-known eigenbasis of the Fourier transform
with eigenvalues $i^n$, the Hermite functions $\psi_n$.
It thus seems enough to simply project on to
Hermite functions of eigenvalue 1. Indeed we 
know that the $\psi_n$ span the space $L^2$ of
square integrable functions. The problem 
is that we also need to allow for spectra that
are not square-integrable. The most obvious examples
of this are compact CFTs, where the spectrum
is given by a sum of Dirac delta functions.

\subsection{Connection to the modular bootstrap}

We now describe the relation between this analysis and the modular bootstrap program of \cite{Hellerman:2009bu,Keller:2012mr,Friedan:2013cba}.
Using modular bootstrap methods, an upper bound
for the dimension of the lowest non-vacuum primary
field of a theory was obtained. 
We are primarily interested in the case of
large $c$, so it is safe to neglect the missing $L_{-1}$ descendants and take the reduced vacuum
character to be $Z_{\tau}(-\sh,-\sh)$.
From (\ref{gC}) we can read off the image
density $g_C$
\be
g_C(\vec u)= \cosh(2\pi u_1\sqrt{2\sh})\cosh (2\pi u_2\sqrt{2\sh})\ .
\ee
We call this the Cardy density, since it is the simplest continuation
of the Cardy regime all the way to $2\sh$.
Clearly $\rho_C$ has a gap of size $2\sh$, since
\be\label{gap}
h+\bar h = \frac{u_1^2+u_2^2}{2}+2\sh \geq 2\sh\ .
\ee
The question is whether we can construct
a density with a larger gap by subtracting a function $g$ from $g_C$.
To put it another way, if can we find a Fourier invariant function $g$ 
such that
\be\label{gcond}
g_C- g = \left\{\begin{array}{cc} 0 &: \vec u\in D\\ \geq 0 &: \vec{u}\in D^c\end{array} \right.
\ee
with the disc $D = \{ u^2 < x^2\}$, then the new density 
$g_C-g$ will have a gap of size $2\sh + \frac{1}{2}x^2$.
Mathematically, the spectrum is a distribution, so we need the (\ref{gcond})
to hold when integrated against test functions. The second line, for instance,
means that for test functions
\be\label{testpositive}
V = \{\varphi(\vec u ) \geq 0 : \vec u \in D^c\}
\ee
we have
\be
\int_{D^c} (g_C-g)\varphi \geq 0\ .
\ee
Define $V^-$ as the space of anti-selfdual even test functions
in $V$, that is functions for which $\hat\varphi = -\varphi$. 
From the condition on $g$ we have that,
for any $\varphi \in V^-$,
\be\label{maincond}
0 = \int_{\mathbb{R}^2} g\varphi = \int_{D}g_C\varphi + \int_{D^c}g\varphi
\leq \int_{\mathbb{R}^2} g_C \varphi\ ,
\ee
where the first equality comes from the fact that $\varphi$ is
anti-selfdual and the last inequality from (\ref{gcond}).
The problem thus reduces to this: Can we construct an anti-selfdual
test function $\varphi$ which is positive outside $D$ 
which contradicts (\ref{maincond}), \ie for which
\be
(g_C,\varphi) < 0 \quad?
\ee 
If we can, then $2\sh +\frac{1}{2}x^2$ is an upper bound for the gap. 
The entire approach thus reduces to finding appropriate
test functions. In practice the main issue is
checking positivity. 

In fact, this approach is equivalent to \cite{Hellerman:2009bu,Friedan:2013cba}:
one way to choose test functions is as linear
combinations of Hermite functions, which are
indeed a eigenbasis of the Fourier transform.
To ensure the positivity condition (\ref{testpositive}),
we need to check positivity of Hermite polynomials.
Those polynomials are the same as the ones 
obtained from differential operators in the
bootstrap literature, and we recover the
same bound, which goes as $4\sh$ for
large $c$. In view of (\ref{gap}), 
one might hope that one could improve this bound using
other families of test functions.  Unfortunately we were not able to do so.

\section{Poincar\'e Series for the partition function}\label{s:Zp}

We now consider the invariance under the
full modular group,
\be
Z(\tau) = Z(\gamma\tau),~~~~~\gamma\in SL(2,\Z)\ .
\ee
For a given state of dimension $(\Delta, \bD)$ we introduce the energy $E$ and angular momentum $J$
\be
E=\D+\bD,~~~~~J = \D-\bD\in \Z\ .
\ee
We will write $\tau = x+iy$.
Let us compute the Poincar\'e series $\Zp_{\D,\bD}(\tau)$ that we obtain
from the contribution of a primary field of weight
$(\D,\bD)$ and its descendants.
We  want to generalize (\ref{Ssum}) to a sum
over the modular group $\SL$
\be\label{PoincareNaive}
\Zp_{\D,\bD}(\tau)=\sum_{\gamma\in \SL} \Zc_{\D,\bD}(\tau|_\gamma)\ , \qquad \tau|_\gamma = \frac{a\tau+b}{c\tau+d}\ .
\ee
The sum (\ref{PoincareNaive}) is divergent for
two reasons. First, since $\J \in \Z$,
$\Zp$ is already invariant under $\tau\mapsto \tau+1$.
It is easy to remove this divergence: we should only sum over $\SL/\Gcusp$, where
$\Gcusp$ is the stabilizer of the cusp at $i\infty$
generated by $T: \tau\mapsto\tau+1$.
The resulting series however is still divergent.
The problem is with the imaginary part of
$\tau|_\gamma$:
\be
\Im(\tau|_\gamma) = \frac{y}{|c\tau+d|^2}\ ,
\ee
which goes to 0 for large $c$ and $d$. 
So the sum diverges like
$\sum |c\tau+d|^{-1}$ and must be regularized. 
A priori there are several possible ways to do so.
We will use the one that is suggested by the above
remarks: the sum over $y^s q^\D \bq^\bD$,
\be\label{ESeries}
E(\tau,s,\D,\bD)=\sum_{\gamma\in \SL/\Gcusp}  \left( y^s q^\D \bq^\bD\right)_\gamma\ ,
\ee
converges if $\Re(s)>1$. Since $E(\tau,s,\D,\bD)$ is
analytic in $s$ in that region, we can try to analytically continue
 to $s=1/2$, and define
this to be the regularization of (\ref{PoincareNaive}), 
\be
\Zp_{\D,\bD}(\tau)=E(\tau,1/2,\D,\bD) \ .
\ee
It was proven in \cite{Maloney:2007ud} (and we will
show  below) that the analytic
continuation of (\ref{ESeries}) is indeed regular
at $s=1/2$, so that this
regularization scheme works.
This regularization scheme is certainly not unique, so one should 
ask whether it is the right one to use for a given
physical problem.
We will return to this in section~\ref{s:gravity}.
For the moment it is enough for us to know that it gives a
well-defined answer which is modular invariant.

Since the result is invariant under $T$ we can write a
Fourier expansion
\be\label{Es}
E(\tau,s,\D,\bD) = y^s q^\D \bq^\bD + \sum_{\j} e^{2 \pi i \j x} E_{\j}(s,\E,\J) \ ,
\ee
where the first term corresponds to the identity element
of $\SL$ and the $E_{\j}$ contain the other images.
The advantage of this approach is that we can find explicit expressions for the $E_{\j}(s,\E,\J)$, which
 allow us to check their physical properties.
This approach was first described in \cite{Maloney:2007ud}, though 
our results will be a bit more detailed. 

First, note that the sum  over $\SL/\Gcusp$ in (\ref{ESeries}) can be written as 
a sum over relatively prime integers $c,d$ with $c\geq0$, or,
defining $d= d'+ \hat \j c$, as a triple sum over $c\geq0, d'\in \Z/c\Z, \hat\j\in\Z$.
Next we can perform a Poisson resummation, which gives a sum over spins $\j$
of the Fourier transform of the summand.
This gives (\ref{Es}), where $E_{\j}$ includes a sum
over $c$ and $d'$. 
This in turn we expand in a power series
in $\E$, summing over $m$,
\be
 E_{\j}(s,\E,\J) = \sum_{m=0}^\infty E_{\j,m}(s,\E,\J) = \sum_{m=0}^\infty I_{\j,m} (s,\E,\J)y^{1-m-s} Z_{\j,\J}(m+s)\ .
\ee
The sum over $c$ and $d'$ has been absorbed in the Kloosterman zeta function 
\be
Z_{\j,\J}(m+s)  = \sum_{c=1}^\infty c^{-2(m+s)} S({\j}, \J;c)\ ,
\ee
which is a sum over Kloosterman sums
\be
S(\j,J;c)=\sum_{d\in (\Z/c\Z)^*} \exp\left\{ 2 \pi i {\j d+ J d^{-1} \over c}\right\}\ .
\ee
Note that $Z$ converges if $s$ is large enough.
Appendix~\ref{s:Kloosterman} contains additional information
about Kloosterman zetas and their analytic continuation.
The Fourier integral
\be\label{Imn}
I_{\j,m} (s,\E,\J)= {(2\pi)^m\over m!} \int_{-\infty}^\infty dT e^{-2\pi i \j T y} (1+T^2)^{-m-s}(-E- i J T)^m\ ,
\ee
also converges for large $s$. The Fourier transform from $T$ to $y$
comes from the Poisson resummation mentioned above.
We can compute $E_{\j,m}(s,\E,\J)$ by evaluating
the integral $I_{\j,m}$ and the Kloosterman $Z$ explicitly for large $s$,
and then continue them analytically
to $s=1/2$ to obtain expressions for 
\be
E_{\j,m}(\E,\J) := E_{\j,m}(1/2,\E,\J)\ .
\ee
We will now discuss the various cases in detail.

\subsection{$\J=0,\j=0$}
Let us first discuss the case $\J=0$ and $\j=0$. From (\ref{Imn}) we obtain
\be
I_{0,m}(s,\E,0) = \frac{ 2^m\pi^{m+1/2}  \Gamma(-\frac{1}{2}+m+s)}{m!\Gamma(m+s)}(-\E)^m\ .
\ee
Note that for the $m=0$ term the $s$ regularization was needed, since
it diverges for $s=1/2$.
The Kloosterman zeta function can be evaluated explicitly as in (\ref{Z00}), giving
\be
E_{0,m}(s,\E,0) =  
{\zeta(2(m+s)-1) \over \zeta(2(m+s))}
\frac{ 2^m\pi^{m+1/2} \Gamma(-\frac{1}{2}+m+s)}{m!\Gamma(m+s)}
(-\E)^m y^{1-m-s} \ .
\ee
We can now take the limit $s\rightarrow\frac{1}{2}$. 
As pointed out above, the only problematic term is $m=0$, where
the integral $I_{0,m}$ diverges. This is cancelled by a zero of the Kloosterman
sum, since $\Gamma(s-1/2)/\zeta(2s) \rightarrow 2$. We find
\begin{eqnarray}
E_{0,0}(\E,0) &=& -y^{1/2}\\
E_{0,m}(\E,0) &=&  
{\zeta(2m) \over \zeta(2m+1))}\frac{ 2^m\pi^{m+1/2}  }{m\Gamma(m+1/2)}
(-\E)^m y^{1/2-m}
\end{eqnarray}
where we have used $\zeta(0)=-1/2$.

\subsection{$\J=0,\j\neq0$}
For the terms with non-vanishing spin $\j$ it is useful to define
\be\label{Ijm}
I_{\j,m}(s)=I_{\j,m}(s,-1,0)=\frac{(2\pi)^m}{m!}\int_{-\infty}^\infty dT e^{-2\pi i\j Ty}(1+T^2)^{-m-s}\ 
\ee
so that $I_{\j,m}(s,\E,0)=(-\E)^mI_{\j,m}(s)$. Since
$\j\neq0$, for any $m\geq0$ the integral converges for $s=1/2$, so there is no need for regularization.
Defining
\be
c_m = \frac{2^{m+1}\pi^{2m+1/2}}{m!\Gamma(m+1/2)}=\frac{2^{3m+1} \pi ^{2 m} }{(2m)!}\ ,
\ee
we obtain
\be
I_{\j,m}(1/2)=c_m  |\j|^{m} y^{m}K_{m}(2\pi y |\j|)\ ,
\ee
where $K_m$ is a modified Bessel function of the second kind.
The Kloosterman zeta
function $Z_{\j,0}$ can be evaluated explicitly (\ref{Z0J}) and 
makes the $m=0$ term vanish. We find
\begin{eqnarray}
E_{\j,0}(\E,0) &=& 0\\ 
E_{\j,m}(\E,0) &=&
\frac{\sigma_{2m}(\j)}{|\j|^{2m}\zeta(2m+1)}c_m|\j|^m(-\E)^m 
y^{1/2}K_{m}(2\pi y |\j|)\ .
\end{eqnarray}

\subsection{$\J\neq 0,\j=0$}
Let us now the case where the original state has spin $\J\neq0$. The terms with
$\j=0$ can again be evaluated explicitly.
For $J\ne0$ and $\j=0$ the integral is
\be\label{IJneq0}
\ba
I_{0,m} (s,E,J)=   {(2\pi)^{m}J^m\over m!\Gamma(m+s)}&\left(\cos\left({m \pi \over 2}\right) \Gamma\left({1+m\over 2}\right)
\Gamma\left({m-1\over2}+s\right)
{}_{2}F_{1}\left({m-1\over2}+s,-{m\over 2};{1\over2};{E^{2}\over J^2}\right)\right. \\ 
&\left.-m\sin\left({m \pi\over 2}\right)\Gamma\left({m\over 2}\right)\Gamma\left({m\over2}+s\right)
{E\over J} {}_{2}F_{1}\left({1-m\over2},{m\over2}+s;{3\over2};{E^{2}\over J^2}\right)\right)\ ,
\ea
\ee
and the Kloosterman zeta gives
\be
 Z_{0,\J}(m+s)= {\sigma_{2(m+s)-1}(\J)\over \J^{2 (m+s)-1}\zeta(2(m+s))}\ .
\ee
For $m=0$ we again need to be careful about divergences. From (\ref{IJneq0}) we obtain
\be
E_{0,0}(s,\E,\J)= {\sigma_{2s-1}(\J)\over \J^{2 s-1}\zeta(2s)}\frac{\sqrt{\pi}\Gamma(s-1/2)}{\Gamma(s)}y^{1-s}
\rightarrow^{s\rightarrow1/2} 2\sigma_{0}(\J)y^{1/2}\ .
\ee
For $m>0$, we can set $s=1/2$ directly to obtain
\be
\ba
I_{0,m} (E,J)=   {2\pi^{m+1/2}J^m\over m\Gamma(m+1/2)}&\left(\cos\left({m \pi \over 2}\right) 
{}_{2}F_{1}\left({m\over2},-{m\over 2};{1\over2};{E^{2}\over J^2}\right)\right. \\ 
&\left.-m\sin\left({m \pi\over 2}\right)
{E\over J} {}_{2}F_{1}\left({1-m\over2},{m\over2}+1/2;{3\over2};{E^{2}\over J^2}\right)\right)\\
=&{2\pi^{m+1/2}J^m\over m\Gamma(m+1/2)}{}_2F_1\left(m,-m,\frac{1}{2};\frac{1+E/J}{2}\right)
\ea
\ee
where in the last line we have used the quadratic transformation
(28) in 2.1.5 of \cite{MR698779}.
We can write this in terms of Chebyshev polynomials of the first
kind $T_n(x)$, 
\be\label{Chebyshev}
T_n(x)= {}_2F_1(-n,n;\frac{1}{2};\frac{1-x}{2}) 
= \frac{(x-\sqrt{x^2-1})^n+(x+\sqrt{x^2-1})^n}{2}
= \cosh(n\cosh^{-1}(x)) \ ,
\ee
where the last representation is only valid for $x \geq 1$.
In total we get
\be
I_{0,m}(E,J) = {2\pi^{m+1/2}J^m\over m\Gamma(m+1/2)}T_m(-\E/\J)\ ,
\ee
which gives
\begin{eqnarray}
E_{0,0}(\E,\J)&=&2\sigma_{0}(\J)y^{1/2}\\
E_{0,m}(\E,\J)&=&{2\pi^{m+1/2}\sigma_{2m}(\J)\over m\Gamma(m+1/2)|\J|^{m}\zeta(2m+1)}T_m(-\E/|\J|)y^{1/2-m}\ ,
\end{eqnarray}
where we have used that $T_m(-x)=(-1)^mT(x)$.

\subsection{$\J\neq0,\j\neq0$}
Using (\ref{Ijm}) we can write
\be
I_{\j,m}(s,\E,\J)=\left(-\E+\J(2\pi \j)^{-1}\frac{d}{dy}\right)^mI_{\j,m}(s)\ .
\ee
This is allowed because for $s\geq1/2$ and $m\geq1$ the integral
and its derivatives converge absolutely.
Again we can set $s=1/2$ without any convergence issues. 
Unfortunately we are no longer able to simplify the Kloosterman zeta
further. Moreover for $m=0$, we need to continue $Z_{\j,\J}(s)$
analytically to $s=1/2$. As we discuss in appendix~\ref{s:Kloosterman}, 
the continuation never has a pole at $s=1/2$, so that this regularization gives
indeed a finite result.
We obtain
\be
E_{\j,m}(\E,\J)= Z_{\j,\J}(m+1/2)c_m|\j|^{m}y^{1/2-m}\left(-\E+\J(2\pi \j)^{-1}\frac{d}{dy}\right)^my^{m}K_{m}(2\pi y |\j|) ~ .
\ee

\section{Inverse Laplace transforms and the spectrum}\label{s:spectrum}

We will now describe the spectrum
of primary states which come from the modular image of a given primary of dimension $(\Delta, \bD)$.  We will write the primary counting partition function $Z^p_{\Delta, \bD}$ coming from the Poincar\'e images of the $(\Delta, \bD)$ state as 
\be
Z_{\D,\bD}(\tau)= |\eta(\tau)|^{-2}
\left(q^\D \bq^\bD + \sum_{\j}\int_0^\infty d\e \rho(\e,\j) e^{2\pi i x \j} e^{-2\pi y \e}\right)\ .
\ee
Here $e$ and $j$ are the energy and angular momentum of the new states coming from Poincar\'e series, while $E$ and $J$ are used to denote the energy and angular momentum of the original ``seed" primary state.  The density of states $\rho(\e,\j)$ is related to $E_{\j}(1/2,\D,\bD)$
by an inverse Laplace transform
\be
y^{-1/2}E_{\j}(1/2,\E,\J) = \int_0^\infty d\e  e^{-2\pi y \e} \rho(\e,\j)\ ,
\ee
or, using the decomposition into $m$,
\be
y^{-1/2}E_{\j,m}(\E,\J) = \int_0^\infty d\e e^{-2\pi y \e} \rho_{\j,m}(e)\ ,
\ee
where
\be
\rho(\e,\j)= \sum_{m=0}^\infty \rho_{\j,m}(e)\ .
\ee
So we need to find the inverse Laplace transform of $y^{-1/2}E_{\j,m}$.
In this section we will explicitly compute $\rho(\e,\j)$
and prove that it satisfies the properties claimed in the introduction.

Our goal is to show that all new primaries obtained from the Poincar\'e series 
satisfy cosmic censorship, that is
\be\label{cc}
\rho(\e,\j) = 0 \qquad \textrm{if}\ |\j| > e
\ee
for any $\E$ and $\J$. This is already visible 
from the asymptotic behavior of the $E_{\j,m}(\E,\J)$.
The Laplace transform of a function
$\rho(\e,\j)$ which vanishes for $\e<\j$
decays as $e^{-2\pi j y}$ for $y\rightarrow \infty$.
The asymptotic behavior of the $E_{\j,m}$ is given by the Bessel
function
\be
E_{\j,m} \sim K_m(2\pi y|j|) \sim e^{-2\pi y |j|}\ ,
\ee
from which we expect that the $\rho_{\j,m}(\e)$ should
satisfy (\ref{cc}). This will be shown explicitly below.

We also want to check positivity, \ie under what condition
a primary contributes only positive terms to 
the spectral density $\rho(e,j)$. This question
is more difficult to answer.
We will show that, after a minor modification of the regularization
scheme, for $-\E$ large enough and $|\J|$ not too big,
the contributions to the spectral density are indeed positive.

\subsection{Spinless primary fields $\J=0$}
For $\j=0$, using 
\be
\LP^{-1}(y^{-m}) = (2\pi)^m \frac{\e^{m-1}}{(m-1)!}
\ee 
the $\rho_{0,m}$ are:
\be \label{rhoJ0j0}
\rho_{0,0} = -\delta(\e)\ , \qquad \rho_{0,m} =2^{m-1}{\zeta(2m) \over \zeta(2m+1))}
c_m(-\E)^m \e^{m-1} \ .
\ee
For $\j\neq0$ we need to find the inverse Laplace transform
of modified Bessel functions. 
It is useful to define the variable $s=2\pi y |\j|$,
and its Laplace dual variable $t=\e/|\j|$. We
then denote the Laplace transform as 
$\tilde h(s) = \LP(h(t))= \int_0^\infty dt h(t)e^{-st}$,
so that $\rho(\e) = |\j|^{-1}\LP^{-1}(y^{-1/2}E_{\j,m})(\e/|\j|)$.
We can then use equation (12) in section 5.15 of \cite{MR0061695}
to write $\LP^{-1}(K_m(s)) =f_m(t)$ with
\be\label{invLPBessel}
 f_{m}(t) = \left\{\begin{array}{cc}
0 &: 0 < t < 1\\
(t^2-1)^{-1/2} \cosh(m \cosh^{-1}(t)) &: 1 < t
\end{array} \right.
\ee
Note that we can also express this in terms of Chebyshev polynomials $T_n(x)$
defined in (\ref{Chebyshev}).
It follows that
\be\label{rhoJ0jn0}
\rho_{\j,0}(\e) = 0\ , \qquad \rho_{\j,m}(\e) = \frac{\sigma_{2m}(\j)}{|\j|^{2m}\zeta(2m+1)}c_m|\j|^{m-1} (-\E)^m 
f_{m}(\e/|\j|)\ .
\ee
Again from the form of $f_m$ it follows that $\rho_{\j,m}$ satisfies
(\ref{cc}). 

\subsection{Positivity}\label{ss:positivity}
Let us now discuss the positivity of the spectrum.
From the expressions above we see that if we choose $\E < 0$,
\ie a primary state in the censored region, 
then all the individual contributions $\rho_{\j,m}$ other than $\rho_{0,0}$ are positive.
Here we have used that both $\Gamma$ and
$\zeta$ are positive for the arguments given,
and that $\sigma_{2m}(j)=\sigma_{2m}(-j) \geq 0$. 
The only problem is $\rho_{0,0}$, which gives
minus a delta distribution at the origin.

We can compensate for this by adding 
a primary field with $\e=0$ and $\j=0$.
A modular invariant way of doing so
is, for instance, to add the partition function
of a free boson compactified on an $S^1$
at the self-dual radius:
\be
Z = \frac{1}{|\eta(\tau)|^2}\sum_{k,l\in\Z}q^{(l+k)^2/4}\bar q^{(l-k)^2/4}\ .
\ee
We know that this is modular invariant
and has manifestly positive coefficients.
Moreover all primary fields in this partition function
 have $\Delta,\bD\geq0$, so they lie
in the uncensored region. The primary $l=0,k=0$
then exactly cancels the delta distribution,
so that the total sum has positive spectrum.

\subsection{Primaries with spin $\J\neq0$}
Let us now turn to primaries with spin.
For $\j=0$ it is again straightforward to invert the Laplace
transform to obtain
\be
\rho_{0,0}=2\sigma_{0}(\J)\delta(\e) \qquad \rho_{0,m}= 
{\sigma_{2m}(\J)\over |\J|^{m}\zeta(2m+1)}c_m T_m(-\E/|\J|)e^{-1+m}\ .
\ee
Clearly this satisfies (\ref{cc}).
Positivity on the other hand depends on the Chebyshev
polynomials $T_m(-\E/|\J|)$.
We can use the fact that $T_m(x) \geq 0$ for $x\geq1$.
It follows that $\rho$ is indeed non-negative as long $-\E \geq |J|$.
If $|J|$ is outside this range the Chebyshev polynomials oscillate
and determining positivity is more subtle.

Next we need to deal with $\j\neq0$. Using again the variables $s=2\pi y |\j|$ and $t=\e/|\j|$,
define
\be\label{DK}
\nu_{\j,m}(\e) = \LP^{-1}\left(s^{-m}\left(-\E+sgn(\j)\J\frac{d}{ds}\right)^ms^{m}K_{m}(s)\right)\ ,
\ee
so that
\be\label{nudef}
\rho_{\j,m}(\e) = Z_{\j,\J}(m+1/2)c_m|\j|^{m-1} \nu_{\j,m}(\e)\ .
\ee
Let us now compute $\nu_{\j,m}$.
Without loss of generality we can take $\j>0$, since otherwise 
we choose $-\J$. 
We define the differential operator $D_s$
\be
s^{-m}(-\E+\J\partial_s)s^m = -\E +\J(\partial_s +ms^{-1}) =: D_s\ ,
\ee
so that we can write (\ref{DK}) as $\nu_{\j,m}=\LP^{-1}(D_s^m K_m(s))$. Next, we  compute the inverse Laplace
transform of $D_s$ by
\be\label{opLP}
\LP^{-1}\left((-\E +\J(\partial_s +ms^{-1})) \tilde h(s)\right) = (-\E -\J t)h(t) +\J m \int_0^t h(t')dt'
=: D_t h(t)\ 
\ee
where $\tilde h$ is the Laplace transform of the function $h$. 
The Laplace transformation has thus turned
the differential operator $D_s$ into a integral operator $D_t$.
We then have
\be\label{nufromD}
\nu_{\j,m}(t)= D_t^m \LP^{-1}(K_m(s)) = D_t^m f_m(t)\ .
\ee
From (\ref{opLP}) we see that if the function $h(t)$
vanishes for $t<1$, so does $D_t h(t)$,
from which it follows that
indeed $\nu_{\j,m}(\e)=0$ for $|\j| > \e$.
This establishes (\ref{cc}).

To check positivity, let us compute $\nu_{\j,m}$
somewhat more explicitly. Since all 
the functions involved vanish for $t<1$,
we can introduce
a new variable $\cosh u = t$ such that 
\be
f_m(u) =\frac{\cosh mu}{\sinh u}\Theta(u)
\ee
where $\Theta(u)$ is the Heaviside step function. In the new
variable, $D_t$ acts as
\be
D_t h(u) = (-\E - \J \cosh u)h(u) + \J m\int_0^u du' \sinh u' h(u')\ .
\ee
We can then evaluate
\begin{multline}
D_t f_k(u) = \left(-\E \frac{\cosh ku}{\sinh u} 
- \frac{\J}{\sinh u}(\cosh u \cosh ku -\frac{m}{k} \sinh u\sinh ku)\right)\Theta(u)\\
= -\E f_k(u) - \J f_{k-1}(u)
-\frac{\J}{2}(1-\frac{m}{k})(f_{k+1}(u) -f_{k-1}(u))
\end{multline}
From this and (\ref{nufromD}) it follows that
\be
\nu_{\j,m}(\e) = \sum_{k=0}^m a_k f_k(\e/|\j|)\ ,
\ee
for some coefficients $a_k$ which are polynomials in $\E$ and $\J$.
To check positivity, we need the first few leading
terms in $-\E$,
\be\label{nuexpand}
\nu_{\j,m}(\e) = (-\E)^mf_m-m(-\E)^{m-1}\J f_{m-1}
+\J^2(-\E)^{m-2}\frac{m}{4}\left((2m-1)f_{m-2}-f_m\right) + O((-\E)^{m-3})\ . 
\ee
This follows from 
\be
D^k_t f_m = (-\E)^k f_m - k\J(-\E)^{k-1}f_{m-1}
+\J^2(-\E)^{k-2}\frac{k(k-1)}{4(m-1)}\left((2m-1)f_{m-2}-f_m\right)+O(\J^3)\ ,
\ee
which can be checked by recursion.

Let us now check that (\ref{nudef}) is positive. There are two separate issues here, 
namely positivity of $\nu_{\j,m}$ and positivity of the Kloosterman zeta $Z_{\j,\J}(m+1/2)$.
For the first note that $f_m(\e/|\j|) \geq f_n(\e/|\j|)$ for $m>n$. 
If we choose $-\E$ much larger than $\J$,
from (\ref{nuexpand}) we conclude that
the leading term dominates,
\be\label{nuleading}
\nu_{\j,m}(\e) \simeq (-\E)^m f_m(\e/|\j|)\ ,
\ee
which we know is positive.
As for the Kloosterman zeta, for $m\geq1$ by the remarks in appendix~\ref{s:Kloosterman} 
we know that $Z_{\j,\J}(m+1/2)$ is always positive. 
For $m=0$ the situation is more subtle since 
we need to continue the Kloosterman zeta analytically.
It is thus conceivable that the term with $m=0$ is negative. 
We can show however that the combined contribution
of the $m=0$ and the $m=2$ term is positive.
For this we use the bound
\be
|Z_{\j,\J}(1/2)| \leq |\j|^2K_\J\ 
\ee
from the appendix.
The total contribution of the two terms
is
\be
\rho_{\j,0}(\e)+\rho_{\j,2}(\e) \geq \frac{1}{2}c_2|\j|\nu_{\j,2}(\e)
-c_0K_\J|\j|\nu_{\j,0}(\e)
\simeq |j|(\frac{1}{2}c_2(-\E)^2 f_2(\e/|\j|)- c_0 K_\J f_0(\e/|\j|))
\ee
which for large enough $-\E$ is positive for all $\j$.

\subsection{The vacuum contribution}
So far we have checked that for $-\E$ large enough,
the contribution of a primary field is positive.
Let us now investigate the contribution of the vacuum.
This is actually
slightly different from the other primaries, due to the vanishing of the $L_{-1}$ descendants.
The vacuum contribution to the primary
partition function is
\be\label{vacuum}
q^{\D}\bar q^{\bD}(1-q-\bar q+ q\bar q)
\ee
with $\D =\bD = -\frac{c-1}{24}=-\sh$. 
Because of the missing $L_{-1}$ descendants
there are potential negative contributions to the
primary spectrum. We will now show that for $c$
large enough, the total spectrum is still positive.
Setting $E=-2\sh$, we need
to sum the contributions
\be
\rho^{vac}(\e,\j) = \rho_{-2\sh,0} - \rho_{-2\sh+1,1}-\rho_{-2\sh+1,-1}+\rho_{-2\sh+2,0}\ .
\ee 
Let us first check the contributions to the $\j=0$ states.
In this case we get
\be\label{sixdelta}
\rho^{vac}_{0,0} = -6\delta(\e)
\ee
and
\be
\rho^{vac}_{0,m} = \left(
\zeta(2m) ((2\sh)^m+(2\sh-2)^m)
-4\cdot2^{-m}T_m(2\sh-1)\right)
\frac{2^{m-1}c_m}{\zeta(2m+1)}\e^{m-1}\ .
\ee
Using $\zeta(2m)>1$ and the fact that
$x^m + (x-2)^m - 2^{-m+2}T_m(x-1) \geq 0$ for $x\geq1$, 
we find that $\rho_{0,m}$ is indeed positive for $\sh \geq\frac{1}{2}$.
In fact for large $\sh$ we expand the bracket
\be
2^{-2m}(2(2\sh)^m-2m(2\sh)^{m-1})+ \frac{1}{2} m (2 m-1)(2\sh)^{m-2}+\ldots
\ee
The sum over the first term is exponentially supressed,
and the second term gives the expected Cardy behavior
$\rho^{vac}(\e,0) \sim \exp(2\pi\sqrt{8\sh\e})$.

Next consider $\j\neq0$. 
From (\ref{rhoJ0jn0}) and (\ref{nudef}), the total contribution is
\begin{multline}\label{vacneg}
\rho^{vac}_{\j,m}(\e) = c_m|\j|^{m-1} \bigl(Z_{\j,0}(m+1/2)\left(\nu_{\j,m}^{-2\sh,0}(\e)+\nu_{\j,m}^{-2\sh+2,0}(\e)\right)\\
-Z_{\j,1}(m+1/2)\nu_{\j,m}^{-2\sh+1,1}(\e)-Z_{\j,-1}(m+1/2)\nu_{\j,m}^{-2\sh+1,-1}(\e)\bigr)\ .
\end{multline}
The situation is more involved than for $\j=0$. In particular,
it is no longer true  that every single term $\rho^{vac}_{\j,m}(\e)$
is positive. We will argue, however, that the sum over $m$ is positive.
The reason this is possible is that, unlike the $\j=0$ case,
for any value of $\e$ the $f_m(\e/|\j|)$ are monotonically growing in $m$. 
So it is possible for terms with large $m$
to dominate the spectrum everywhere. 
For $\j=0$ this argument would have failed since $\e^{m-1}$
only grows monotonically if $\e>1$, so terms
with small $m$ can dominate for small $\e$.

For $\sh$ large, due to the prefactor $c_m$, the sum will peak at $m \sim \sh^{1/2}$.
For large $m$ we have $Z_{\j,\J}(m+1/2) = 1+ O(4^{-m-1/2})$,
so at the peak of the sum the subleading terms are suppressed
exponentially as $\sim 4^{-\sh^{1/2}}$. We will thus only keep
the leading term.
Using (\ref{nuexpand}) we find that the first two
terms cancel, and the third term is positive,
\begin{multline}
\rho^{vac}_{\j,m}(\e) = c_m|\j|^{m-1} \bigl(\left(2\sh)^m - 2m(2\sh)^{m-1} +2m(m-1)(2\sh)^{m-2} \right)f_m\\
+(-2(2\sh)^m+2m(2\sh)^{m-1}-m(m-1)(2\sh)^{m-2})f_m
-(2\sh)^{m-2}\frac{m}{2}((2m-1)f_{m-2}-f_m)\bigr)\\
=  c_m|\j|^{m-1}(2\sh)^{m-2} m(m-1/2)(f_m-f_{m-2})+O(\sh^{m-3})\ .
\end{multline}
Since $f_m > f_{m-2}$ this is indeed positive.

\subsection{Asymptotic behavior and comparison to Cardy}
Finally let us compare our results with the usual
Cardy results. 
We have 
\be\label{Cardy}
\rho_{\Delta,\bD} \sim \exp (2\pi\sqrt{-4\Delta\Delta'})\exp(2\pi\sqrt{-4\bD\bD'})
=\exp(2\pi\sqrt{-(\E+\J)(\e+\j)})\exp(2\pi\sqrt{-(\E-\J)(\e-\j)})
\ee
For spinless primaries we we can
approximate the Poincar\'e series expression for $-Ee$ large enough as
\be
\rho(\e,0) = \sum_{m=1}^\infty \rho_{0,m} \simeq (2\e)^{-1}\exp(2\pi\sqrt{-4\E\e})\ ,
\ee
which indeed agrees with (\ref{Cardy}).
This also works for $\j\neq0$ but $|\j|\ll\e$, where 
we can evaluate (\ref{rhoJ0jn0}) using 
$f_m(t) \sim (2t)^{m-1}$ and $\sigma_{2m}(\j) \sim \j^{2m}$
for large $m$ to get
\be
\rho(\e,\j) \sim \sum_m \frac{2^{3m+1}\pi^{2m}}{(2m)!}|\j|^{m-1}f_m(\e/|\j|)
\sim \exp\left(2\pi \sqrt{-2\E}\sqrt{\e+\sqrt{\e^2-\j^2}}\right) \ 
\ee
which agrees with (\ref{Cardy}).

Note that for a generic CFT, the Cardy behavior becomes
valid only for $\D \gg c$. In our case
for large $c$ the behavior actually becomes valid for 
$\D,\bD \sim 1$ already. Not surprisingly, our partition
functions with a minimal censored spectrum
give an extreme example of the extension of the
Cardy regime discussed in \cite{Hartman:2014oaa}.

\section{Pure gravity and Farey tails}\label{s:gravity}

The Poincar\'e
sum has a physical interpretation in AdS$_3$ gravity. It is 
the sum over all the saddle points of the
classical Euclidean action. Each saddle is a Euclidean continuation of a particular BTZ black hole.
In the case of pure gravity -- gravity without any additional degrees of freedom -- the full partition
function should thus be given by
the Poincar\'e series of (\ref{vacuum})  \cite{Maloney:2007ud}.\footnote{This statement relies on the argument of \cite{Maloney:2007ud} that (\ref{vacuum}) gives the full contribution to the partition function of thermal AdS, at all orders in perturbation theory in $1/c$.  In other words, the perturbative partition function is one-loop exact. The other terms in the Poincar\'e series give non-perturbative (instanton) contributions.}
This sum is divergent and must be regularized.
One possible regularization was presented in section 3.
Let us denote the answer so obtained by $\ZpureP$.
As was pointed out in \cite{Maloney:2007ud}, $\ZpureP$
cannot be the partition function of a healthy dual CFT for two reasons.  First,
it has a continuous spectrum.  Second, it has
a negative density of states at $\e=0$.
However, the regularization scheme we have chosen
is not unique.
In this section we will ask whether there is another physically sensible way to regularize
the sum which gives a different answer, $\Zpure$, which
does not have these problems.

We begin by noting that there 
is another very natural approach
to regularizing the sum, already discussed in \cite{Maloney:2007ud}.
Consider the Laplace operator on $\H$,
\be
\Laplace = -y^2(\partial_y^2+\partial_x^2) = \Im(\tau)^2\partial_\tau\partial_{\bar\tau}\ .
\ee
It is straightforward to show that $\Delta$ is invariant under 
$SL(2,\Z)$. Moreover the Poincar\'e series $E(\tau, s, \Delta, \bD)$ satisfies the following recursion relation in $s$,
\be\label{Erec}
(\Laplace - s(1-s))E(\tau,s,\Delta,\bD) = -2\pi (\Delta+\bD) sE(\tau,s+1,\Delta,\bD)
+(2\pi)^2\Delta\bD E(\tau,s+2,\Delta,\bD)\ .
\ee
If $s$ is such that $\Laplace -s(1-s)$ is invertible,
that is if $\lambda=s(1-s)$ is not in its spectrum,
then we can use (\ref{Erec}) to define the analytic
continuation recursively: Starting out
with $s$ such that the right hand side converges, we obtain
$E(s)$ by successively applying $(\Laplace - s(1-s))^{-1}$. 
The success of this procedure thus depends on the
spectrum of $\Laplace$, which in turn depends on
space of functions on which we define its action.
We review some features of the spectral theory of $\Laplace$ in appendix~\ref{s:Laplace}.
It turns out that the spectrum is discrete on
the space $\cL^2$ of square integrable functions on $\H$.
For such functions this recipe for analytic continuation
gives something finite and unique away from a discrete set of points in
the $s$-plane. 
For $\D,\bD<0$, however, $E(\tau,s,\D,\bD) \notin \cL^2$.
For such functions it turns out
that the Laplacian is not invertible for any value of $s$,
so that this regularization scheme does not give a unique answer.

To proceed, let us discuss physically what
properties we should require of our regularized partition function $\Zpure$.
We certainly want it to be modular
invariant. Since we are considering pure gravity,
we also want no new censored states in the spectrum.
This implies that $\Zpure-\ZpureP$ does not grow
exponentially for $y\rightarrow\infty$.
Note that we cannot exclude polynomial
growth here.
Finally note that
\be\label{Laplace}
(\Laplace-1/4)\Zpure
\ee
is an actual physical observable which gives a finite
result without any regularization.  It is the expectation value of the stress tensor, integrated over the torus. We thus require 
that $(\Laplace-1/4)(\Zpure-\ZpureP)=0$. 
This, together with the behavior at $y\rightarrow\infty$, implies that $\Zpure$
differs from $\ZpureP$ at most by
 a \emph{Maass form}. As we discuss in appendix~\ref{s:Laplace},
the only Maass form
with eigenvalue $1/4$ is the Eisenstein
series $\hat E(\tau,1/2)$.  The addition of this term to $\ZpureP$, however, is not enough
to compensate the negative term (\ref{sixdelta}),
much less to make the spectrum discrete.
This shows that there is no way 
to regularize the Poincar\'e series which 
gives a healthy pure gravity partition function $\Zpure$.

It is important to note, however, that $\rho_{pure}$ only fails to be physical
at subleading order in $c$.  The leading ${O}(c)$ behavior is perfectly fine.  
As noted in section~\ref{ss:positivity}, we can easily
remove the negative term (\ref{sixdelta}) by adding
an $O(1)$ number of primary fields. 
Likewise, by shifting the weights of all
states by $O(1)$, one could obtain a modular invariant partition
function with discrete spectrum.
We conclude that it is possible to
find a partition function $\Zpurep$
which is modular invariant, has a positive
discrete spectrum, no uncensored
states other than the vacuum descendants,
and which differs from the Poincar\'e series
$\ZpureP$ only by terms which are subleading in the large central charge (i.e. bulk semi-classical) limit
\be
\Zpurep  \xrightarrow[c\to\infty]{}
\ZpureP + O(1)\ .
\ee
The existence of such partition functions is a central result of this paper.

These $O(1)$ terms should be regarded  as intrinsically quantum mechanical contributions to the sum over geometries.\footnote{Similar correction terms appear when one tries to study the partition function of pure gravity at small central charge \cite{Castro:2011zq}.  In this case, deviations from the semi-classical (large $c$) results appear because the Virasoro representations develop null states at small $c$.  It seems unlikely that a similar phenomenon could be responsible for the $O(1)$ contributions arising at large $c$, however.}  They are distinct from the saddle point contributions to the partition function, which have the feature that they are dominant in some region of moduli space and exponentially subleading in other regions of moduli space. The new $O(1)$ pieces are sub-dominant everywhere in moduli space, and -- since they are finite in the large $c$ limit -- cannot be interpreted as  contributions from semi-classical saddles.  They might, for example, come from the contribution to the path integral of a new saddle point with Planckian curvature. 
Unfortunately, we not know how to study such saddles, nor do we know of a principle which would allow one to determine the $O(1)$ pieces uniquely.  
We also note that the existence of this $\Zpurep$  does not guarantee the
existence of a corresponding CFT. 
Should such a CFT exist, however, it could be interpreted as the holographic dual of pure AdS$_3$ gravity in the semi-classical limit.

\bigskip

\textbf{Acknowledgments} We thank Daniel Friedan, Tom Hartman, Arnaud Lepage-Jutier, Daniel Whalen and Edward Witten 
for useful discussions. CAK is supported by the Rutgers
New High Energy Theory Center and by U.S. DOE Grants No.~DOE-SC0010008,
DOE-ARRA-SC0003883 and DOE-DE-SC0007897. CAK would like to thank the
Harvard University High Energy Theory Group for hospitality.

\appendix

\section{Harmonic analysis}\label{s:Laplace}
\subsection{Mass forms}
We will follow \cite{MR1942691,MR3100414}. 
Let $\mathbb{H}$ be the upper half plane such
that $\Im(\tau)>0$. It has a natural $SL(2,\Z)$
invariant metric
\be
ds^2 = y^{-2}(dx^2+dy^2)\ .
\ee
A modular
function is a function $f: \H \rightarrow \C$
such that
\be
f(\gamma \tau) = f(\tau) \qquad \forall \gamma \in SL(2,\Z)\ .
\ee
Let us call the space of such functions $\cA$.
There is a natural inner product on the space of
modular functions,
\be
\langle f,g\rangle = \int_{F} f(\tau)\bar g(\tau)y^{-2}dxdy
\ee
where the integral is over a fundamental region $F$.
The space of functions $\mathcal{L}^2$ of square
integrable functions is a Hilbert space.
Unfortunately most of the functions we consider are not
in $\mathcal{L}^2$. Clearly any function with a polar
part diverges as $y \rightarrow \infty$. However,
even when we eliminate the polar part there is no
guarantee that the resulting function
will be square-integrable due to the prefactor
$y^{1/2}$ coming from the descendants. The contribution
of the a constant function for instance diverges
logarithmically.

Nevertheless let us proceed with the analysis.
The Laplace operator 
\be
\Laplace = -y^2(\partial_y^2+\partial_x^2) = -\Im(\tau)^2\partial_\tau\partial_{\bar\tau}\ 
\ee 
is invariant under
$SL(2,\Z)$, and symmetric with respect
to the inner product.
A function $f\in\cA$ which is an eigenfunction
of the Laplace operator
\be
(\Laplace-\lambda)f=0\ , \qquad \lambda = s(1-s) \ ,
\ee
is called an automorphic Maass form. Let us
denote by $\cA_s$ the space of Maass forms
with eigenvalue $\lambda=s(1-s)$.
Ultimately the goal is thus
to decompose modular functions into
Maass forms. Let us thus analyze
the eigenfunctions of $\Laplace$.

\subsection{Eisenstein series}
One class of such eigenfunctions can be constructed from
Eisenstein series.
Take $\psi$ a smooth function
on $\R^+$. We then consider
the Poincar\'e series
\be
E(\tau|\psi) = \sum_\gamma \psi(\Im(\gamma \tau))\ ,
\ee
which converges absolutely if
\be
\psi(y) \ll y (\log y)^{-2} \qquad y\rightarrow0\ .
\ee
If we choose $\psi(y) =y^s$ with $\Re(s)>1$ then we obtain
the Eisenstein series
\be
E(\tau,s) = \sum_\gamma (\Im(\gamma \tau))^s\ .
\ee
For $\Re(s) < 1$ we can use analytic continuation.
Clearly $E(\tau,s)$ is an eigenfunction of $\Laplace$
of eigenvalue $s(1-s)$, \ie it is a Maass form.
It is however not square integrable. 
For our purposes it is actually more useful to consider the functions 
\cite{MR633666}
\be
\hat E(\tau,s) = (2\Lambda(1/2))^{-1}\Lambda(s)E(\tau,s)\ ,\qquad \Lambda(s)=\pi^{-s}\Gamma(s)\zeta(2s)\ ,
\ee
which are clearly still in $\cA_s$. They are regular
except for simple poles at $s=0$ and $s=1$, and
their Fourier expansions have the explicit expressions
\be
2\Lambda(1/2)\hat E(\tau,s) = \Lambda(s)y^s + \Lambda(1-s)y^{1-s}
+2y^{1/2}\sum_{j=1}^\infty j^{s-1/2}\sigma_{1-2s}(j)K_{s-1/2}(2\pi ijy)\cos(2\pi jx)\ .
\ee
In particular for $s=1/2$ we have
\be
\hat E(\tau,1/2) = (\gamma-\log(4\pi))y^{1/2}+
y^{1/2}\sum_{j=1}^\infty \sigma_{0}(j)K_{0}(2\pi jy)(e^{2\pi i jx} + e^{-2\pi i jx})\ 
\ee
where we have used that $2\Lambda(1/2) = \gamma -\log(4\pi)$ 
with $\gamma$ the Euler constant.

We are looking for a spectral decomposition of the space $\cL^2$,
that is square integrable modular functions. To this end
it is useful to define various subspaces. First
define $\cB$ the space of smooth bounded modular
functions, which is dense in $\cL^2$.
Next consider Eisenstein series where $\psi$ is 
compactly supported in $\R^+$, so that
$E(\tau|\psi)$ is bounded on $\H$ and hence
in $\cB$. We call such an $E(\tau|\psi)$
an \emph{incomplete Eisenstein series}, and denote
their space $\cE$. We have the inclusion
\be
\cE \subset \cB \subset \cL^2 \subset \cA\ .
\ee
Next let us consider the orthogonal complement 
of $\cE$ in $\cB$. Any modular function
$f\in\cA$ can be decomposed as
\be
f(\tau) = \sum_n f_n(y) e^{2\pi i n x}\ .
\ee
Denote by $\cC$ the space of all smooth
bounded modular functions for which
$f_0(y)=0$. If in addition $f$ is a Mass form, then
we call it a \emph{cusp form}.
It turns out that
\be
\cL^2 = \overline{\cC}\oplus \overline{\cE}\ ,
\ee
where the bar stands for the closure.

We have $\Laplace :\cC \rightarrow \cC$ and
$\Laplace : \cE \rightarrow \cE$.
It turns out that $\Laplace$ has pure point spectrum
in $\cC$, \ie $\cC$ is spanned by cusp forms, 
whereas on $\cE$ the eigenpacket of the continuous
spectrum is spanned by the Eisenstein series $E(\tau,s)$
analytically continued to $\Re(s) =1/2$, and some
point spectrum on the segment $1/2<s\leq 1$.
This means that for any modular function $f\in\cL^2$
we have the spectral decomposition 
\be\label{spectral}
f(\tau) = \sum_j \langle f, u_j\rangle u_j(\tau) + 
\frac{1}{4\pi}\int_{-\infty}^\infty dr \langle f, E(\cdot,1/2+ir)\rangle E(\tau,1/2+ir)\ ,
\ee
where the $u_j(\tau)$ are eigenfunctions of $\Laplace$
with discrete eigenvalues $\lambda_j$.
It turns out in the case at hand with $\SL$ 
$1/4$ is not in the point spectrum of $\cC$,
so that $\hat E(\tau,1/2)$ is indeed the only
Maass form of eigenvalue $1/4$.

\section{Kloosterman Sums}\label{s:Kloosterman}
Define the Kloosterman zeta as
\be
Z_{\j,\J}(m+s)  = \sum_{c=1}^\infty c^{-2(m+s)} S({\j}, \J;c)\ ,
\ee
which is a sum over Kloosterman sums
\be
S(\j,J;c)=\sum_{d\in (\Z/c\Z)^*} \exp\left\{ 2 \pi i {\j d+ J d^{-1} \over c}\right\}\ .
\ee
Clearly we have $Z_{\j,\J}(s)=Z_{\J,\j}(s)$.
In some special cases we can evaluate $Z$ explicitly,
namely
\footnote{Recall the divisor function $\sigma_x(n)=\sum_{d|n} d^x$, from which
it follows $\sigma_{-x}(n) = n^{-x}\sigma_{x}(n)$.}
\be\label{Z0J}
Z_{0,\J}(m+s)= \sum_{c=1}^\infty c^{-2(m+s)} S(0, \J;c)= {\sigma_{2(m+s)-1}(\J)\over \J^{2 (m+s)-1}\zeta(2(m+s))}
\ee
and 
\be\label{Z00}
Z_{0,0}(m+s)= {\zeta(2(m+s)-1) \over \zeta(2(m+s))}\ .
\ee
There are a few estimates for $Z_{\j,\J}(m+s)$.
Note that $S(\j,\J,1)=1$, and that trivially
$|S(\j,\J,c)| \leq c$. For $s=1/2$ 
and $m\geq1$ we can thus estimate
\be
|Z_{\j,\J}(m+1/2) -1 | \leq \sum_{c=2}^\infty c^{-2m} < \int_1^\infty dc c^{-2m} = \frac{1}{2m-1}\ , 
\ee
from which in particular it follows that
$Z_{\j,\J}(m+1/2)$ is positive for $m \geq 1$.

\subsection{Analytic continuation of the Kloosterman zeta}
We will now use the spectral theory presented in appendix~{s:Laplace} 
to analytically continue $Z(s)$ to $s=1/2$. 
In this we follow Selberg's original paper \cite{MR0182610}
and also \cite{MR689644,MR1763900}. 
For $m>0$ let us define the auxiliary Poincar\'e series
\be
P_m(\tau,s) = \sum_{\gamma\in\Gamma/\Gamma_\infty} e^{2\pi i m \gamma(\tau)} \frac{y^s}{|c\tau+d|^{2s}}\ .
\ee
Like the ordinary Eisenstein series, this converges for 
$\Re(s)>1$. 
When expanding this series, we will encounter Kloosterman
zeta functions just as we did in the analysis. We can
thus read off the analytic continuation of $Z(s)$ from
the analytic continuation of $P_m$.
Crucially, because $m>0$, $P_m$ is square integrable,
unlike the Poincar\'e series we encountered in the main
body of this article. (To deal
with the case $m<0$, following a remark in \cite{MR1763900},
we take instead the function $\overline{P_m(\tau,s)}$, which
is again in $\cL^2$.)
Analogous to (\ref{Erec}) we have the recursion relation
\be\label{Prec}
P_m(\tau,s) = 4\pi m s R_{s(1-s)}(P_m(\tau,s+1))\ 
\ee
for $\Re(s)>1$, where
\be
R_{s(1-s)}= (\Laplace-s(1-s))^{-1}\ .
\ee
This means that unless $s$ happens to lead
to an eigenvalue $\lambda=s(1-s)$ of the Laplacian,
we can take (\ref{Prec}) to define the analytic
continuation of $P_m(\tau,s)$ to $\Re(s) \leq 1$.
To do this in practice, we use the spectral decomposition
(\ref{spectral}) into eigenfunctions of $\Laplace$,
on which the resolvent $R_{s(1-s)}$
acts by simple multiplication.

Note that unlike the case discussed  in section~\ref{s:gravity}
this works because this time
we can restrict to $\cL^2$ functions, for
which the spectral decomposition (\ref{spectral})
makes sense. Let us first discuss the cusp
part of $P_m$, \ie the part of the decomposition
coming from the discrete part of the spectrum.
The general idea is that for $\H/\SL$, 
$1/2$ is not a discrete eigenvalue of the Laplacian,
so that the contribution from the cusp part is
regular. In a more detailed computation, \cite{MR689644} 
use this fact to provide upper bounds in the more
general case for 
the analytically continued $Z_{\j,\J}(s)$, but they are
not valid for $s=1/2$. The reason for this is that 
for a general Fuchsian group $s=1/2$ can 
indeed be a discrete eigenvalue of the Laplacian. 
We can easily repeat their
analysis, in particular keeping track of the explicit $\j$
dependence of their bound. In the following we will
use the usual $O(\cdot)$ notation with the understanding
that the implied constant will never depend
on any of the variables in the expression.

First let us evaluate Lemma 1 in \cite{MR689644}. We have
\be\label{Sarnak1}
\int_F |P_m(\tau,1/2)|^2 \frac{dx dy}{y^2}
\leq 4\pi^2m^2|R_{1/4}|^2 \int_F |P_m(\tau,3/2)|^2 \frac{dx dy}{y^2}
= O(m^2)
\ee
where we have used that integral converges,
and that the smallest eigenvalue discrete
of $\Laplace$ on $\H/SL(2,\Z)$ is of order $\lambda_1 \simeq 90$ so that
$|R_{1/4}| \leq |1/4-90|^{-1}$.
To apply Lemma 2, we need to bound $R(s)$.
We have 
\be
|R_{m,n}(1/2,c)|\leq\int_0^\infty\int_{-\infty}^\infty \frac{y^2}{(x^2+1)^{1/2}}
\left|\exp(-2\pi im\frac{x-i}{yc^2(x^2+1)})-1\right| e^{-2\pi ny} \frac{dxdy}{y}
\ee
The square of the absolute value we evaluate as the sum of
\be 
\sin^2 2\pi m \frac{x}{yc^2(x^2+1)}\exp(-\frac{2\pi m}{yc^2(x^2+1)}) 
\leq \left( \frac{2\pi m x}{yc^2(x^2+1)}\right)^2
\ee
and
\begin{multline}
\left((1-\cos \frac{2\pi m x}{yc^2(x^2+1)})\exp(-\frac{2\pi m}{yc^2(x^2+1)})+
(1-\exp(-\frac{2\pi m}{yc^2(x^2+1)}))\right)^2\\
\leq \left(  \frac{1}{2}\left( \frac{2\pi m x}{yc^2(x^2+1)}\right)^2
+ \frac{2\pi m}{yc^2(x^2+1)})\right)^2
\end{multline}
In total we can thus bound
$|R_{m,n}(1/2,c)| = O(c^{-2}m^2)$,
which together with $Z_{m,n}(3/2)\leq 1$ yields
\be\label{Sarnak2}
R(1/2)=O(m^2)\ .
\ee
Combining (\ref{Sarnak1}) and (\ref{Sarnak2}) with
Lemma 2 then gives
\be\label{Zbound}
Z_{m,n}(1/2) = O(n^2m^2)\ .
\ee
This in particular shows that the contribution from the cusp part
is regular at $s=1/2$. As we pointed above however,
we also need to worry about the contributions from the
continuous part of the spectrum. A more detailed 
argument \cite{MR1942691} shows however that the
contribution at $s=1/2$ of the continuous part of
the spectrum vanishes, so that 
$Z(1/2)$ is indeed regular. We thus conjecture that 
(\ref{Zbound}) continues to hold when one takes
into account the continuous spectrum.
(Note that for our general argument to hold, 
a weaker bound is sufficient:
It is enough for $Z_{m,n}$ to only
grow polynomially in $m$.)

\bibliographystyle{../ytphys}
\bibliography{../ref}

\def\cprime{$'$}
\providecommand{\href}[2]{#2}\begingroup\raggedright\begin{thebibliography}{10}

\bibitem{Cardy:1986ie}
J.~L. Cardy, ``{Operator Content of Two-Dimensional Conformally Invariant
  Theories},''
\href{http://dx.doi.org/10.1016/0550-3213(86)90552-3}{{\em Nucl.Phys.}
  {\bfseries B270} (1986) 186--204}.

\bibitem{Carlip:2000nv}
S.~Carlip, ``{Logarithmic corrections to black hole entropy from the Cardy
  formula},'' \href{http://dx.doi.org/10.1088/0264-9381/17/20/302}{{\em
  Class.Quant.Grav.} {\bfseries 17} (2000) 4175--4186},
\href{http://arxiv.org/abs/gr-qc/0005017}{{\ttfamily arXiv:gr-qc/0005017
  [gr-qc]}}.

\bibitem{Hellerman:2009bu}
S.~Hellerman, ``{A Universal Inequality for CFT and Quantum Gravity},''
  \href{http://dx.doi.org/10.1007/JHEP08(2011)130}{{\em JHEP} {\bfseries 1108}
  (2011) 130},
\href{http://arxiv.org/abs/0902.2790}{{\ttfamily arXiv:0902.2790 [hep-th]}}.

\bibitem{Qualls:2013eha}
J.~D. Qualls and A.~Shapere, ``{Bounds on Operator Dimensions in 2D Conformal
  Field Theories},''
\href{http://arxiv.org/abs/1312.0038}{{\ttfamily arXiv:1312.0038 [hep-th]}}.

\bibitem{Hartman:2014oaa}
T.~Hartman, C.~A. Keller, and B.~Stoica, ``{Universal Spectrum of 2d Conformal
  Field Theory in the Large c Limit},''
\href{http://arxiv.org/abs/1405.5137}{{\ttfamily arXiv:1405.5137 [hep-th]}}.

\bibitem{Maldacena:1997re}
J.~M. Maldacena, ``{The Large N limit of superconformal field theories and
  supergravity},'' {\em Adv.Theor.Math.Phys.} {\bfseries 2} (1998) 231--252,
\href{http://arxiv.org/abs/hep-th/9711200}{{\ttfamily arXiv:hep-th/9711200
  [hep-th]}}.

\bibitem{Dijkgraaf:2000fq}
R.~Dijkgraaf, J.~M. Maldacena, G.~W. Moore, and E.~P. Verlinde, ``{A Black hole
  Farey tail},''
\href{http://arxiv.org/abs/hep-th/0005003}{{\ttfamily arXiv:hep-th/0005003
  [hep-th]}}.

\bibitem{Banados:1992wn}
M.~Banados, C.~Teitelboim, and J.~Zanelli, ``{The Black hole in
  three-dimensional space-time},''
  \href{http://dx.doi.org/10.1103/PhysRevLett.69.1849}{{\em Phys.Rev.Lett.}
  {\bfseries 69} (1992) 1849--1851},
\href{http://arxiv.org/abs/hep-th/9204099}{{\ttfamily arXiv:hep-th/9204099
  [hep-th]}}.

\bibitem{Maldacena:1998bw}
J.~M. Maldacena and A.~Strominger, ``{AdS(3) black holes and a stringy
  exclusion principle},''
  \href{http://dx.doi.org/10.1088/1126-6708/1998/12/005}{{\em JHEP} {\bfseries
  9812} (1998) 005},
\href{http://arxiv.org/abs/hep-th/9804085}{{\ttfamily arXiv:hep-th/9804085
  [hep-th]}}.

\bibitem{Kraus:2006wn}
P.~Kraus, ``{Lectures on black holes and the AdS(3) / CFT(2) correspondence},''
  {\em Lect.Notes Phys.} {\bfseries 755} (2008) 193--247,
\href{http://arxiv.org/abs/hep-th/0609074}{{\ttfamily arXiv:hep-th/0609074
  [hep-th]}}.

\bibitem{Manschot:2007ha}
J.~Manschot and G.~W. Moore, ``{A Modern Farey Tail},''
  \href{http://dx.doi.org/10.4310/CNTP.2010.v4.n1.a3}{{\em
  Commun.Num.Theor.Phys.} {\bfseries 4} (2010) 103--159},
\href{http://arxiv.org/abs/0712.0573}{{\ttfamily arXiv:0712.0573 [hep-th]}}.

\bibitem{Maloney:2007ud}
A.~Maloney and E.~Witten, ``{Quantum Gravity Partition Functions in Three
  Dimensions},'' \href{http://dx.doi.org/10.1007/JHEP02(2010)029}{{\em JHEP}
  {\bfseries 1002} (2010) 029},
\href{http://arxiv.org/abs/0712.0155}{{\ttfamily arXiv:0712.0155 [hep-th]}}.

\bibitem{Keller:2012mr}
C.~A. Keller and H.~Ooguri, ``{Modular Constraints on Calabi-Yau
  Compactifications},'' \href{http://dx.doi.org/10.1007/s00220-013-1797-8}{{\em
  Commun.Math.Phys.} {\bfseries 324} (2013) 107--127},
\href{http://arxiv.org/abs/1209.4649}{{\ttfamily arXiv:1209.4649 [hep-th]}}.

\bibitem{Friedan:2013cba}
D.~Friedan and C.~A. Keller, ``{Constraints on 2d CFT partition functions},''
  \href{http://dx.doi.org/10.1007/JHEP10(2013)180}{{\em JHEP} {\bfseries 1310}
  (2013) 180},
\href{http://arxiv.org/abs/1307.6562}{{\ttfamily arXiv:1307.6562 [hep-th]}}.

\bibitem{Witten:2007kt}
E.~Witten, ``{Three-Dimensional Gravity Revisited},''
\href{http://arxiv.org/abs/0706.3359}{{\ttfamily arXiv:0706.3359 [hep-th]}}.

\bibitem{Friedan:2012jk}
D.~Friedan, A.~Konechny, and C.~Schmidt-Colinet, ``{Lower bound on the entropy
  of boundaries and junctions in 1+1d quantum critical systems},''
  \href{http://dx.doi.org/10.1103/PhysRevLett.109.140401}{{\em Phys.Rev.Lett.}
  {\bfseries 109} (2012) 140401},
\href{http://arxiv.org/abs/1206.5395}{{\ttfamily arXiv:1206.5395 [hep-th]}}.

\bibitem{MR698779}
A.~Erd{\'e}lyi, W.~Magnus, F.~Oberhettinger, and F.~G. Tricomi, {\em Higher
  transcendental functions. {V}ol. {I}}.
\newblock Robert E. Krieger Publishing Co., Inc., Melbourne, Fla., 1981.
\newblock Based on notes left by Harry Bateman, With a preface by Mina Rees,
  With a foreword by E. C. Watson, Reprint of the 1953 original.

\bibitem{MR0061695}
A.~Erd{\'e}lyi, W.~Magnus, F.~Oberhettinger, and F.~G. Tricomi, {\em Tables of
  integral transforms. {V}ol. {I}}.
\newblock McGraw-Hill Book Company, Inc., New York-Toronto-London, 1954.
\newblock Based, in part, on notes left by Harry Bateman.

\bibitem{Castro:2011zq}
A.~Castro, M.~R. Gaberdiel, T.~Hartman, A.~Maloney, and R.~Volpato, ``{The
  Gravity Dual of the Ising Model},''
  \href{http://dx.doi.org/10.1103/PhysRevD.85.024032}{{\em Phys.Rev.}
  {\bfseries D85} (2012) 024032},
\href{http://arxiv.org/abs/1111.1987}{{\ttfamily arXiv:1111.1987 [hep-th]}}.

\bibitem{MR1942691}
H.~Iwaniec, {\em Spectral methods of automorphic forms}, vol.~53 of {\em
  Graduate Studies in Mathematics}.
\newblock American Mathematical Society, Providence, RI; Revista Matem\'atica
  Iberoamericana, Madrid, second~ed., 2002.

\bibitem{MR3100414}
A.~Terras, \href{http://dx.doi.org/10.1007/978-1-4614-7972-7}{{\em Harmonic
  analysis on symmetric spaces---{E}uclidean space, the sphere, and the
  {P}oincar\'e upper half-plane}}.
\newblock Springer, New York, second~ed., 2013.
\newblock \url{http://dx.doi.org/10.1007/978-1-4614-7972-7}.

\bibitem{MR633666}
D.~Zagier, ``Eisenstein series and the {R}iemann zeta function,'' in {\em
  Automorphic forms, representation theory and arithmetic ({B}ombay, 1979)},
  vol.~10 of {\em Tata Inst. Fund. Res. Studies in Math.}, pp.~275--301.
\newblock Tata Inst. Fundamental Res., Bombay, 1981.

\bibitem{MR0182610}
A.~Selberg, ``On the estimation of {F}ourier coefficients of modular forms,''
  in {\em Proc. {S}ympos. {P}ure {M}ath., {V}ol. {VIII}}, pp.~1--15.
\newblock Amer. Math. Soc., Providence, R.I., 1965.

\bibitem{MR689644}
D.~Goldfeld and P.~Sarnak, ``Sums of {K}loosterman sums,''
  \href{http://dx.doi.org/10.1007/BF01389098}{{\em Invent. Math.} {\bfseries
  71} no.~2, (1983) 243--250}.

\bibitem{MR1763900}
W.~d.~A. Pribitkin, ``A generalization of the {G}oldfeld-{S}arnak estimate on
  {S}elberg's {K}loosterman zeta-function,''
  \href{http://dx.doi.org/10.1515/form.2000.014}{{\em Forum Math.} {\bfseries
  12} no.~4, (2000) 449--459}.

\end{thebibliography}\endgroup

\end{document}